\documentclass[12pt]{article}
\pdfoutput=1

\usepackage{color}
\usepackage{amsmath}
\usepackage{amsfonts}
\usepackage{amssymb}
\usepackage{caption}
\usepackage{graphicx}
\usepackage{slashed}            % for slashed characters in math mode
\usepackage{subfig}             % for sub figures
\usepackage{xspace}				% For spacing after commands
\usepackage{cite}
\usepackage[english]{babel}
\usepackage{fancyhdr}
\usepackage{amsmath}
\usepackage{amssymb}
\usepackage{amsfonts}
\usepackage{psfrag}
\usepackage[applemac]{inputenc}
\usepackage{graphicx}
\usepackage{braket}
\usepackage{breqn}
\usepackage{feynmp}
\usepackage{feynmp-auto}
\usepackage{subfig}
\usepackage[export]{adjustbox}
\usepackage[
	colorlinks=true,
	citecolor=black,
	linkcolor=black,
	urlcolor=blue,
	hypertexnames=false]{hyperref}

\def\beq{\begin{eqnarray}}
\def\eeq{\end{eqnarray}}
\def\bean{\begin{equation*}}
\def\eean{\end{equation*}}
\newcommand{\arXiv}[2]{\href{http://arxiv.org/pdf/#1}{{\tt #2/#1}}}
\newcommand{\arXivold}[1]{\href{http://arxiv.org/pdf/#1}{{\tt #1}}}

\numberwithin{equation}{section} 

\newcommand{\MeijerG}[7]{G \begin{smallmatrix} #1 & #2 \\ #3 & #4 \end{smallmatrix} \left( \begin{smallmatrix} #5 \\ #6 \end{smallmatrix} \middle\vert #7 \right) }

\newcommand{\Tr}{\mathrm{Tr~}}

\newcommand{\beqar}{\begin{eqnarray}}
\newcommand{\eeqar}{\end{eqnarray}}

\newcommand{\bsp}{\begin{split}}
\newcommand{\esp}{\end{split}}

\newcommand{\bag}{\begin{align}}
\newcommand{\eag}{\end{align}}

\newcommand{\eV}{\,\mathrm{eV}}
\newcommand{\keV}{\,\mathrm{KeV}}
\newcommand{\MeV}{\,\mathrm{MeV}}
\newcommand{\GeV}{\,\mathrm{GeV}}
\newcommand{\TeV}{\,\mathrm{TeV}}

\newcommand{\sech}{\,\mathrm{sech}}

\definecolor{MyBlue}{rgb}{0.1,0.1,0.8}

\begin{document}
\begin{titlepage}

\begin{center}

	{	% TITLE GOES HERE 	
		\LARGE \bf 
		The  Radion as a Dark Matter Candidate
	}
	
\end{center}
	\vskip .3cm

\begin{center} % AUTHORS HERE
{\bf \  Fayez Abu-Ajamieh\footnote{\tt
		 \href{mailto:abuajamieh@ucdavis.edu}{abuajamieh@ucdavis.edu}
		 }
		} 
\end{center}

\begin{center} 

 % PLACES HERE
	{\it Department of Physics, University of California Davis\\One Shields Ave., Davis, CA 95616}

\end{center}

%%%%%%%%%%%%%%%%%%%%%%%%%%%%%%%%%%%%%%%%%%%%%%%%%%%%%%%%%%%%

\centerline{\large\bf Abstract}

\begin{quote}
I study a class of Randall-Sundrum (RS) models with Spontaneous Breaking of Scale Invariance (SBSI). This class of models implements the  Contino-Pomarol-Rattazzi (CPR) mechanism to achieve SBSI through the small running of an external close-to-marginal scale-breaking operator that leads to a light dilaton/radion with  couplings to matter suppressed by the small running. I show that for radion masses $\lesssim 1$ KeV, it can serve as a Dark Matter (DM) candidate, with a lifetime longer than the age of the universe, and show that the experimental bounds from LHC, Non-Newtonian Gravity and Axion-Like Particle (ALP) searches allow for the existence of such a radion. In spite of the small relic abundance of the light radion produced in this model, we show that it could be possible to obtain the required abundance through additional assumptions, an issue we postpone to the future.

\end{quote}

\end{titlepage}

%%%%%%%%%%%%%%%%%%%%%%%%%%%%%%%%%%%%%%%%%%%%%%%%%%%%%%%%%%
%%%%%%%%%%%%%%%%%%%%%%%%%%%%%%%%%%%%%%%%%%%%%%%%%%%%%%%%%%

\section{Introduction}\label{Chap:Intro}
The Randall-Sundrum (RS) \cite{Randall:1999ee} model provides an attractive solution to the hierarchy problem through the warped geometry of a single extra dimension. In this model, 2 3-branes, the UV and the IR branes, are connected by an extra dimension compactified on $S^{1}$, with the exponential warping of the extra dimension providing the hierarchy between the Planck scale and the Electroweak (EW) scale. The 5D metric has the form:
\begin{equation}\label{eq:Metric}
ds^{2} = e^{-2 A(y)}\eta_{\mu\nu} dx^{\mu}dx^{\nu} - dy^{2}
\end{equation}
where $A$ is the warp factor which depends only on the extra dimension's coordinate $y$. The UV brane is placed at $y=y_{0}$ and the IR brane is placed at $y=y_{1}$ with $y_{0}$ and $y_{1}$ identified, i.e. with $Z_{2}$ symmetry assumed. This metric describes an $AdS_{5}$ space with a negative 5D cosmological constant. The Goldberger-Wise (GW) mechanism \cite{Goldberger:1999uk} stabilizes the size of the extra dimension by introducing a bulk scalar whose negative potential and positive kinetic energy form an effective potential with a stable minimum that defines the size of the extra dimension, which is tuned to yield the required hierarchy between the Planck scale and the EW scale. Fluctuations of the extra dimension around the potential minimum are parametrized by a scalar field, the radion. 

In addition to providing a simple solution to the hierarchy problem, the RS model presents interesting phenomenology, including proposed Dark Matter (DM) candidates. It has been widely suggested in literature that some Kaluza-Klein (KK) eigenmodes of the fields propagating in the bulk might be stable. In particular, models with KK parity \cite{Agashe:2007jb} suggest that the Lightest KK Particle (LKP) is stable, making it an ideal candidate for DM. \cite{Agashe:2004ci, Agashe:2004bm} present warped Grand Unified Theory (GUT) models that have stable KK fermions which can serve as DM candidates. In \cite{Medina:2010mu}, it was proposed that the radion's first excited KK mode could serve as a DM candidate, by assuming that is has odd KK-parity that prevents it from decaying to the ground state. Here I will show that for a certain class of models, the radion itself can be made stable enough to be a DM candidate.\footnote{Radions are closely related to branons, both of which essentially parametrize the fluctuations of the brane into the bulk. Although branons and radions are physically the same, their parametrization in literature is different. For a single extra dimension compactified on $S^{1}$ and parametrized by $Y^{M} = (x^{\mu},Y(x))$, branons are expressed as the brane's pull-back $g_{\mu\nu} = \partial_{\mu} Y^{M} \partial_{\nu} Y^{N} G_{MN}$, which leads to their having even parity and thereby can only appear in a process in pairs, making them stable and therefore a natural DM candidate (see \cite{Baer:2014eja} for a review. Also see \cite{Sundrum:1998sj, Dobado:2000gr, Cembranos:2004pj} for more information on branons, branon DM and their relation to radions). In this paper, a more general parametrization where the radion can have either even or odd parity is allowed, and thus can decay to SM particles (c.f. \ref{eq:MetricFluctuations}).} In particular, I will investigate RS models that provide a natural mechanism for achieving Spontaneous Breaking of Scale Invariance (SBSI).

In general, conformal theories either do not break scale invariance, or have scale invariance broken at an arbitrary scale corresponding to a flat direction \cite{Fubini:1976jm}. The reason for this is that scale invariance leads to an unsuppressed non-derivative self-interaction quartic term for the dilaton, the Pseudo-Nambu-Goldston Boson (PNGB) of scale invariance. The Contino-Pomarol-Rattazzi (CPR) mechanism \cite{CPR, Coradeschi:2013gda, Bellazzini:2013fga} overcomes this by introducing an explicit close-to-marginal scale-breaking bulk operator whose slow running leads to the vanishing of the quartic. One would start with a large quartic in the UV region that has a slow running (i.e. a small $\beta$-function), and then allow for the scale-breaking operator to grow through its small running until the dilaton potential almost vanishes, signaling SBSI.
More specifically, SBSI allows for a potential of the form:
\begin{equation}\label{eq:SBSIquartic}
V_{eff} = F \chi^{4}
\end{equation}
where $\chi$ is a dimensionless field which parametrizes a non-linear realization of the dilaton:
\begin{equation}
\chi = e^{\sigma/\Lambda}
\end{equation}

$F$ must be tuned to $0$ in order for SBSI to occur, which corresponds to tuning the contribution to the 4D cosmological constant exactly to $0$. This does not happen classically. Instead, in the CPR mechanism, an external close-to-marginal operator with dimension $4-\epsilon$~\cite{CPR,Bellazzini:2013fga} explicitly breaks scale invariance. The slow running of the scale-breaking operator
\begin{equation}\label{eq:BetaFn}
\frac{d\lambda}{d\log \mu} =\beta(\lambda) \sim \epsilon \ll 1
\end{equation}
leads to a non-trivial minimum of $V_{eff}$ where the potential almost vanishes and where SBSI occurs. This minimum is determined by the condition:
\begin{equation}\label{SBSIbreaking}
\frac{dF(\lambda(\chi))}{d\chi} \chi + 4 F(\lambda(\chi))=0
\end{equation} 

In effect, the mass of the corresponding dilaton, which is identified with the radion, will be suppressed by the smallness of the running. In addition, the radion's coupling to matter will be suppressed by $\epsilon$ as will be shown later in the paper, leading to a light and potentially stable radion that could serve as a DM candidate, as long as the scale-breaking operator is kept very close to marginal, i.e. as long as $\epsilon \ll 1$. This means that the radion couplings to SM matter (including massless gauge bosons) can be lowered to values $\ll O(\TeV^{-1})$, in contrast to usual RS models where the coupling is generally $\sim O(\TeV^{-1})$. More explicitly, in typical RS models, the radion mass and couplings to the IR brane are related to the logarithm of the hierarchy between the Planck scale and the EW scale, i.e. $\log{\frac{M_{Pl}}{\mu_{EW}}}$, which would make their typical values $\sim O(\TeV^{-1})$, however, with the CPR mechanism, one can achieve much smaller radion masses and couplings that are well below the TeV scale, and without affecting the hierarchy between the Planck scale the EW scale. In summary, the CPR mechanism enables the suppression of the radion's coupling to SM matter, which would make the radion stable enough to serve as a DM candidate. It is this regime of small couplings that we are interested in this paper.\footnote{Notice that if the radion's coupling to SM matter can be suppressed through any means other than the CPR mechanism, it would also serve as a DM candidate as well, however, I am unaware of any other mechanism that can achieve this.}

Another aspect of the CPR mechanism is that the potential at its minimum, which represents the part of the cosmological constant that corresponds to the scale-breaking phase transition, will be suppressed by $\epsilon$ as well. This means that the largest contribution to the cosmological constant can be significantly reduced as well, and even be made consistent with the observed value.

This paper is organized as follows: In Section \ref{Chap:ModelRev}, I review a realization of the CPR mechanism, showing how the radion mass and effective potential are calculated. In Section \ref{Chap:Pheno}, I present some phenomenological aspects of the model. In Section \ref{Chap:Gravity} I will investigate the associated gravity, including the KK gravitons and the contribution to Non-Newtonian gravity. I relegate much of the technical details to Appendix \ref{App:NonNewtonian}. In Section \ref{Chap:Lifetimes}, I summarize the radion's coupling to matter, and calculate the decay widths and lifetimes. I will show that for a considerable part of the parameter space, the radion's lifetime is longer than the age of universe, rendering it stable. In Section \ref{Chap:Bounds} I will discuss the various experimental bounds relevant to the model, and show that the existence of a light DM radion of mass $\lesssim 1$ KeV is allowed by experiment. In Section  \ref{Chap:RelicAbundance} I will discuss the production of a DM radion via freeze-in and calculate the corresponding relic abundance, with much of the details relegated to Appendix \ref{App:RelicAbundance}. In Section \ref{Chap:CosmConst} I discuss the contribution of the radion's effective potential to the cosmological constant and show that it can be made consistent with the observed value, and finally I present the conclusions in Section \ref{Chap:Conclusions}.

\section{Review of the Model} \label{Chap:ModelRev}
I begin with a quick review of the model in \cite{Bellazzini:2013fga}. A holographic realization of the CPR mechanism can be achieved through the action:
\begin{equation}\label{eq:Action}
S =\int d^{5}x \sqrt{g} \Big(-\frac{1}{2\kappa^{2}} \mathcal{R} +\frac{1}{2}g^{MN}\partial_{M}\phi \partial_{N}\phi \hspace{1 mm} - \hspace{1 mm} V(\phi)  \Big)\ -\sum_{i=0,1} \int d^{4}x \sqrt{g_{i}}V_{i}(\phi).
\end{equation}
where $\phi$ is the bulk scalar, $\kappa^{2}$ is the 5D Newton constant related to the 5D Planck scale via $\kappa^{2} = \frac{1}{2M_{5}^{3}}$, and $V_{i}$ are the brane-localized potentials. The the bulk potential is given by:
\begin{equation}
V(\phi) = -\frac{6k^{2}}{\kappa^{2}} -2\epsilon k^{2} \phi^{2}
\end{equation}
with $\epsilon$ representing the small running of the external operator, and $k$ being a scale factor. The running will lead to the formation of a condensate which signals SBSI. The brane-localized potentials are chosen to be:
\begin{equation}\label{eq:BranePotential}
V_{i} = \Lambda_{i} + \lambda_{i} (\phi-v_{i})^{2}
\end{equation}
however, throughout this paper I will only consider the limit $\lambda_{i} \rightarrow \infty$, such that the brane potentials are given by the constant pieces $\Lambda_{i}$. Using the metric in (\ref{eq:Metric}), the Einstein equations and the equation of motion of the bulk scalar can be solved approximately, yielding the solutions:
\begin{equation}\label{eq:Solutions1}
A(y) = -\frac{1}{4}\log{\Big[\frac{\sinh{(4k(y_{c}-y))}}{\sinh{(4k y_{c})}}\Big]}
\end{equation}
\begin{equation}\label{eq:Solutions2}
\phi(y) = v_{0} e^{\epsilon k (y-y_{0})} -\frac{\sqrt{3}}{2\kappa} \log{(\tanh{(2k(y_{c}-y))})}
\end{equation}
where $y_{c} > y_{1}$ corresponds to the location of the condensate formed by the running of the scale-breaking operator. Plugging the solutions in the action and integrating out the extra dimension, the effective brane potentials are found to be:
\begin{equation}\label{eq:UVpotential}
V_{UV} =\mu_{0}^{4} \Big[\Lambda_{0} - \frac{6k}{\kappa^{2}}   \Big]
\end{equation}
\begin{equation}\label{eq:IRpotential}
V_{IR} = \chi^{4} \Big[ \Lambda_{1} +\frac{6k}{\kappa^{2}} \cosh{\Big( \frac{2\kappa}{\sqrt{3}}(v_{1}-v_{0}(\mu_{0}/ \chi)^{\epsilon}) \Big)} \Big] \sech^{2}{ \Big( \frac{\kappa}{\sqrt{3}}(v_{1}-v_{0}(\mu_{0}/ \chi)^{\epsilon}) \Big)}
\end{equation}
where $\mu_{0} = e^{A(y_{0})}$ and $\chi = e^{A(y_{1})}$ is the dilaton field. The UV potential must be tuned to zero in order to obtain a 4D flat space, whereas the IR potential has a small minimum which determines the size of the extra dimension. The minimum of the IR potential represents the contribution to the cosmological constant, and can be found by taking the derivative of (\ref{eq:IRpotential}) with respect to $\chi$. This gives to the leading order in $\epsilon$:
\begin{equation}\label{eq:VIRmin}
V_{IR}^{min} = -\epsilon \frac{2\sqrt{3}k v_{0}}{\kappa} \tanh{\Big(\frac{\kappa}{\sqrt{3}}(v_{1} - v_{0}(\mu_{0}/\chi)^{\epsilon})\Big)} \Braket{\chi}^{4} (\mu_{0}/\chi)^{\epsilon}
\end{equation}

We can see that the potential is proportional to $\epsilon$, which means that the contribution to the cosmological constant is suppressed by the smallness of the running. The potential minimum is assumed to be the origin of the hierarchy between the Planck scale and the EW scale. More specifically, we set:
\begin{equation}\label{eq:HierarchyOrigin}
 k \Braket{\chi} = \zeta \times 1 \hspace{1mm} \TeV
\end{equation}

This will define the scale factor $k$, and here I introduced the factor $\zeta \sim O($a few$)$ which sets the exact location of the hierarchy between 1 and 10 TeV. I will discuss the contribution to the cosmological constant in more detail in Section \ref{Chap:CosmConst}.
The radion's mass and wavefunction are found by using the following ansatz\footnote{This parametrization is more general than the usual treatement of branons. Notice that if we used (\ref{eq:MetricFluctuations}) to find the brane's pull-back, then we would arrive at the same results.} to express the fluctuations in the metric and bulk scalar \cite{Csaki:2000zn}:
\begin{equation}\label{eq:ScalarFluctuations}
\phi(x,y) =\phi_{0} +\varphi(x,y)
\end{equation}
\begin{equation}\label{eq:MetricFluctuations}
ds^{2} = e^{-2A-2F(x,y)}dx^{2} - (1+2F(x,y))^{2}dy^{2}
\end{equation}
where $\phi_{0}$ represents the unperturbed bulk scalar solution given by (\ref{eq:Solutions2}) and $F(x,y)$ is the 5D radion wavefunction. Eq. (\ref{eq:ScalarFluctuations})  and (\ref{eq:MetricFluctuations}) can be used to find Einstein's equations and the EOM of the scalar, which in turn can be solved to find the EOM of the radion. Here I will simply quote the final EOM of $F$ and refer the reader to \cite{Csaki:2000zn} for the full details: 
\begin{equation}\label{eq:RadionEOM}
F'' - 2A' F' - 4A'' F - 2 \frac{\phi_{0}''}{\phi_{0}'}F' + 4A' \frac{\phi_{0}''}{\phi_{0}'} F = -m^{2} e^{2A} F
\end{equation}
subject to the boundary conditions
\begin{equation}\label{eq:RadionBCs}
\big[F' - 2A' F \big]_{y_{0,1}} = 0
\end{equation}
where the prime indicates a deriviative with respect to $y$. This equation can be solved numerically to find the radion's mass and wavefunction. It was shown in \cite{Bellazzini:2013fga} that the radion/dilaton mass to the leading order in $\epsilon$ is given by\footnote{The radion's kinetic term used for calculating the mass in (\ref{eq:DilatonMass}) is unnormalize.}
\begin{equation}\label{eq:DilatonMass}
m_{\sigma}^{2} = \epsilon \frac{32 \sqrt{3}k v_{0}}{\kappa} \tanh \Bigg(  \frac{\kappa}{\sqrt{3}} (v_{1} - v_{0} (\mu_{0}/\chi)^{\epsilon}) \Bigg) \braket{\chi}^{2} (\mu_{0}/\chi)^{\epsilon}
\end{equation} 
which is confirmed by the numerical solution of (\ref{eq:RadionEOM}). Thus the mass of the radion can be made light by keeping the running small. We shall see later that there is virtually no lower limit on the radion's mass, as long as $\epsilon \ll 1$.

\section{Phenomenological Aspects of the Model}\label{Chap:Pheno}
In this section, I will review the main characteristics of the model and calculate the radion mass. The model has many parameters, namely: $\Lambda_{1}$, $v_{0}$, $v_{1}$, $\epsilon$, $\kappa$ and $\zeta$. In order for the model to be viable, it has to produce the hierarchy between the Planck scale and the EW scale as stated in eq. (\ref{eq:HierarchyOrigin}), in addition to reproducing the Planck scale itself:
\begin{equation}\label{eq:RedPlanck}
\overline{M}_{Pl}^{2} = \frac{1}{\kappa^{2}} \int_{y_{0}}^{y_{1}} dy e^{-2A(y)} \approx 2.4 \times 10^{18} \hspace{1 mm} \GeV
\end{equation}
where the the integral goes over the entire orbifold. If we fix $v_{1}$ and $\kappa$ in units of $k$ and keep $\epsilon$ and $\zeta$ as free parameters, then the two requirements fix $v_{0}$ and $\Lambda_{1}$. Throughout this paper, I set $\kappa = 10^{-2}$ and $v_{1} = 4.5$ in units of $k$, which fixes $\Lambda_{1} = -60022.5$, and tune $v_{0}$ against $\epsilon$ to keep the potential minimum (and thus the hierarchy and size of the extra dimension) fixed. The parameter $\zeta$ is chosen to have values between 1 and 10 and has the function of exactly localizing the scale of SBSI between 1 and 10 TeV. It also serves to determine the masses of the KK modes and avoid the lower bounds set by LHC searches. Note that $M_{Pl}$ is insensitive to $\epsilon$, while it is linear in $\zeta$, therefore in order to keep the Planck mass within acceptable limits, while avoiding the bounds from LHC searches, $\zeta$ should be chosen to have a value of a few.

\begin{figure}[!tbp] 
  \centering
  \begin{minipage}[b]{0.45\textwidth}
    \includegraphics[width=\textwidth]{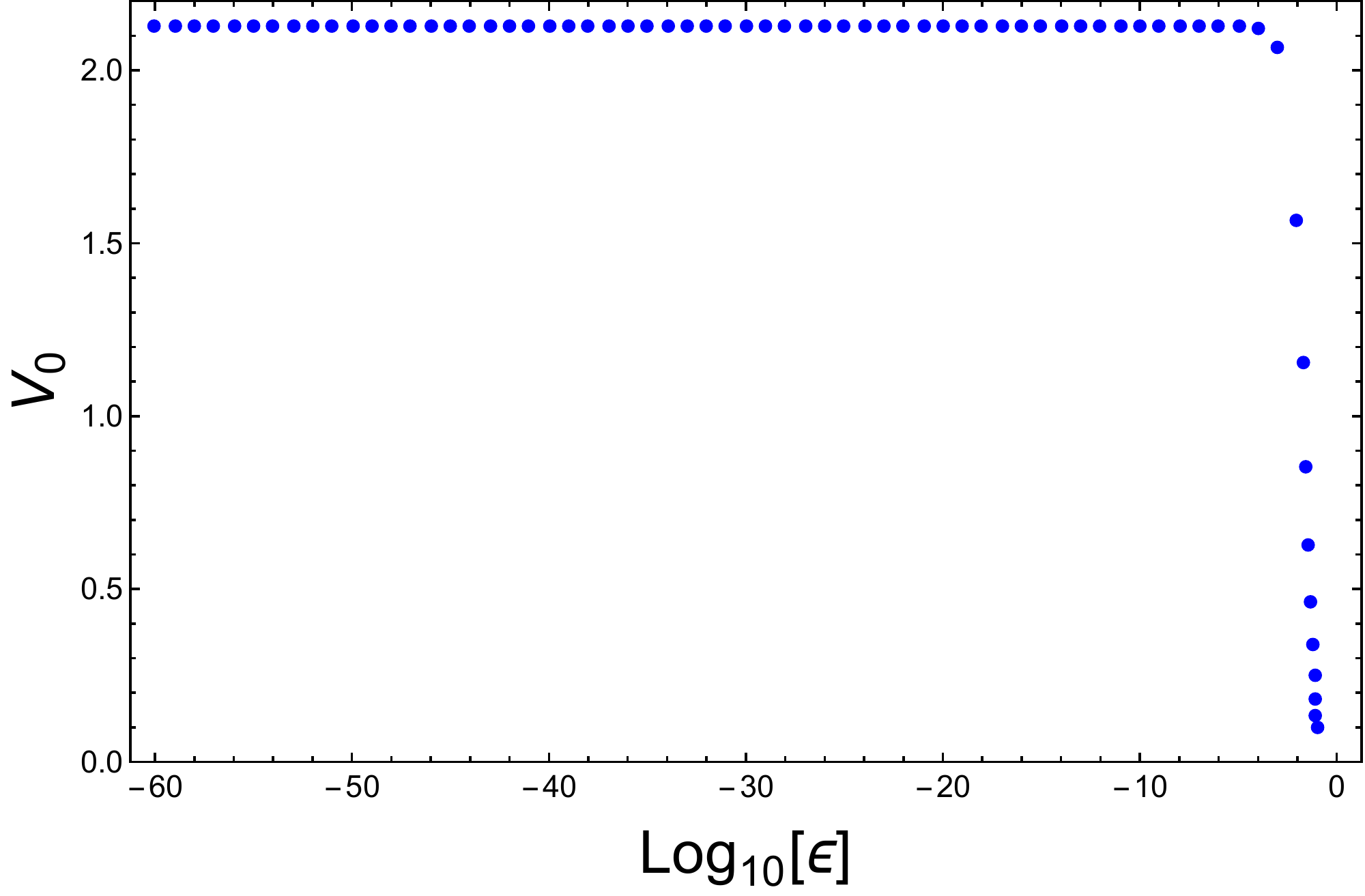}
  \end{minipage}
  \hfill
  \begin{minipage}[b]{0.48\textwidth}
    \includegraphics[width=\textwidth]{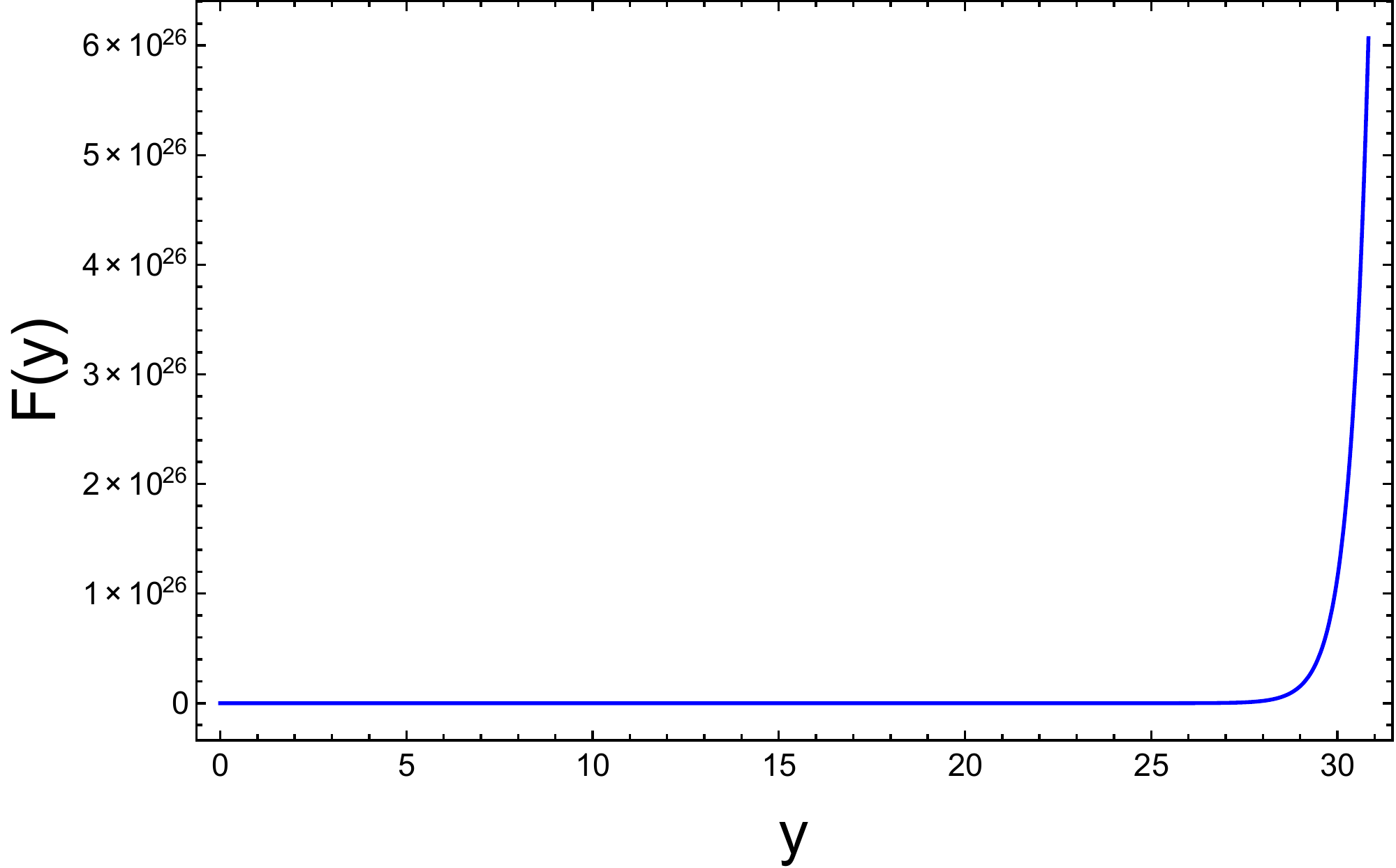}
  \end{minipage}
      \caption{(Left): UV brane potential $v_{0}$ Vs $\log_{10}{\epsilon}$, (Right): Radion wavefunction profile along the extra dimension.}
         \label{fig:V0andF0}
\end{figure}

The left panel of Fig. \ref{fig:V0andF0} shows $v_{0}$ Vs. $\log_{10}{\epsilon}$. The plot shows that the UV potential increases when $\epsilon$ becomes smaller, then saturates to $\sim 2.1$. The reason for this is most obvious if one looks at (\ref{eq:VIRmin}): When $\epsilon$ becomes smaller, the running gets slower and the growth in the quartic potential associated with the scale-breaking operator becomes smaller, so in order to keep the minimum and the hierarchy fixed, one has to increase the potential of the UV brane. When $\epsilon$ becomes very close to zero, $(\mu_{0}/\chi)^{\epsilon} \approx 1$, and the minimum becomes essentially independent of  $v_{0}$.

\begin{figure}[!ht]
 \centering
  \includegraphics[height=.3\textheight]{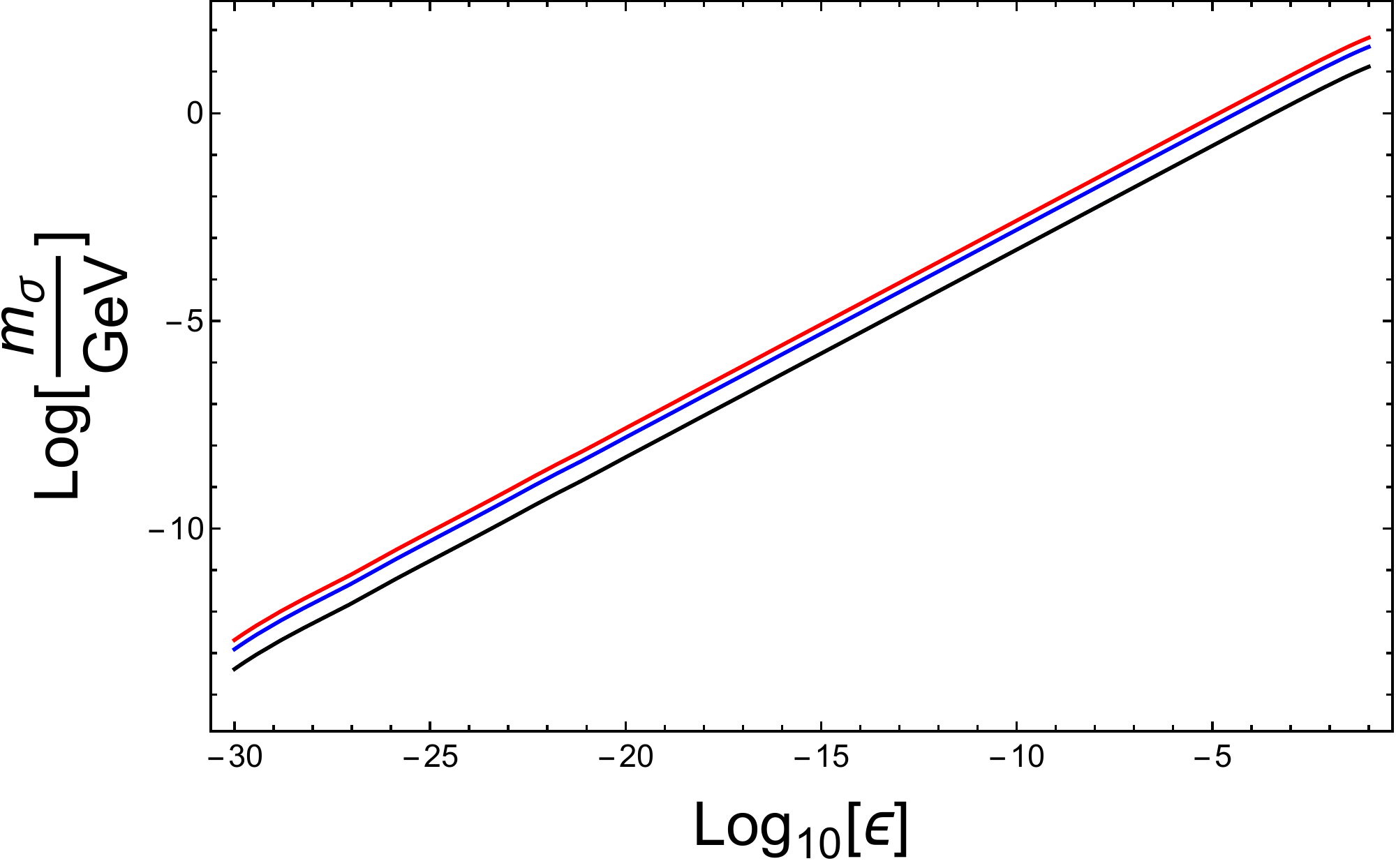}
  \caption{$\log_{10}{m_{\sigma}}$ Vs. $\log_{10}{\epsilon}$ for $\zeta = 1$ (black), $\zeta = 3$ (blue) and $\zeta = 5$ (red).}
  \label{fig:alphaDened}
\end{figure}

Next I turn to calculating the radion mass and wavefunction. Using the solutions (\ref{eq:Solutions1}) and (\ref{eq:Solutions2}), equation (\ref{eq:RadionEOM}) and the boundary condition (\ref{eq:RadionBCs}) can be solved numerically. The right panel of Fig. \ref{fig:V0andF0} shows the radion wavefunction. As shown by the plot, the radion wavefunction is strongly peaked near the IR brane, which means that the radion's coupling to matter localized on the UV brane will be strongly suppressed compared to matter localized on the IR brane. 
The radion mass as a function of $\epsilon$ is shown in Fig. \ref{fig:alphaDened}. As expected, the mass (squared) of the radion is linear in $\epsilon$ to leading order, which confirms (\ref{eq:DilatonMass}). By lowering $\epsilon$, one can reduce the mass of the radion from its natural value of $\sim O($TeV$)$ as in the original RS model, to below eV scale. Note that $\zeta$ has only a small effect on the radion's mass.

\section{Gravity Sector}\label{Chap:Gravity}
In this section, I will calculate the masses of the KK graviton tower and compare them with the latest LHC searches. I will also investigate the modification to Newtonian gravity due to these KK modes and due to the radion, in addition to finding the overall gravity potential.
\subsection{EOM and KK Tower}
I begin by finding the graviton's KK tower. Expanding the fluctuations around the Minkowski space:
\begin{equation}\label{eq:GravitonMetric}
ds^{2} = e^{-2A(y)} \big( \eta_{\mu\nu} + h_{\mu\nu}(x,y) \big)dx^{\mu}dx^{\nu} -dy^{2}
\end{equation}
and using the KK decomposition:
\begin{equation}\label{eq:KKdecomposition}
h_{\mu\nu}(x,y) = \tilde{h}_{\mu\nu}(x) \Psi_{n}(y)
\end{equation}
the EOM and boundary conditions are found to be \cite{Goldberger:1999wh,Davoudiasl:1999jd,Das:2015zxa}:
\begin{equation}\label{eq:GravitonEOM}
\Psi_{n}'' - 4A' \Psi_{n}' +m_{n}^{2} e^{2A} \Psi_{n} = 0
\end{equation}
\begin{equation}\label{eq:GravitonBCs}
\big[ \Psi_{n}' +2 A' \Psi_{n}  \big]_{y_{0,1}} =0
\end{equation}

Notice here that $m_{n} = 0$ is always a solution for the EOM. This zero mode solution describes the massless graviton, and it has a constant profile across the bulk. On the other hand, the profiles of the massive KK gravitons are peaked near the IR brane. This reflects the known fact that the coupling of the zero mode graviton to IR matter is suppressed by the Planck scale, while the couplings of massive KK modes are of $O(\TeV^{-1})$. More specifically, the coupling term of massive gravitons is given by
\begin{equation}\label{eq:GravitonCoupling}
\kappa T_{\mu\nu} \Psi_{n}(y_{i}) h^{\mu\nu}(x)
\end{equation}

\begin{figure}[!ht]
 \centering
  \includegraphics[height=.3\textheight]{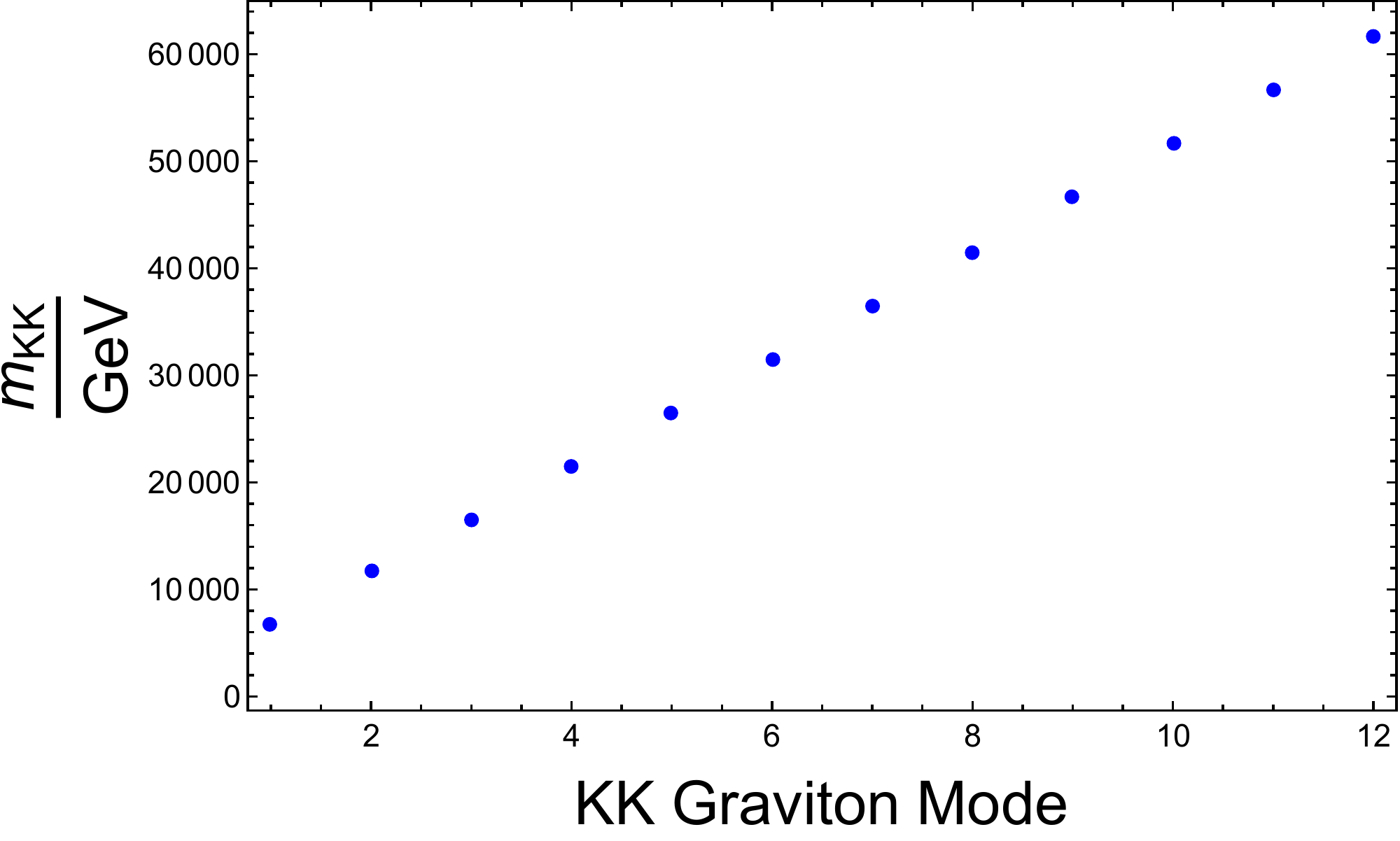}
  \caption{The masses of the first few KK graviton modes for $\zeta =1.6$.}
  \label{fig:KKGravitons}
\end{figure}

where $T_{\mu\nu}$ is the stress energy tensor, and $y_{0}(y_{1})$ gives the gravitons' coupling to UV(IR) matter. Using the warp factor in (\ref{eq:Solutions1}), the EOM and boundary conditions can be solved numerically in order to find the masses of the KK gravitons. The mass of the first KK graviton is $4218 \zeta$ GeV, while the latest LHC searches \cite{Aaboud:2016tru} indicate the masses below $\sim 5000$ GeV are excluded, which constrains $\zeta$ to be $\gtrsim 1.2$. The first few KK masses are plotted in Fig. \ref{fig:KKGravitons}. As the plot shows that the relationship is fairly linear and can be well-fitted with the equation:
\begin{equation}\label{eq:KKGravitons}
m_{n} \approx 3224 \hspace{1mm}  \zeta \hspace{1mm} n \hspace{1mm}  \GeV \hspace{2 mm} (n > 1)
\end{equation}

An important consequence of this result is that the LHC searches can be easily avoided with the appropriate choice of $\zeta$. For instance, if we set $\zeta \sim 3.1$, we can easily push the mass of the first excited graviton all the way to $\gtrsim 13$ TeV, which makes it beyond the reach of the LHC. The same applies for the KK masses of all bulk fields, which extends the viability of the RS model beyond the reach of the LHC, while maintaining the main feature of the model, namely providing a solution to the hierarchy problem. Notice however, that there is an upper limit on $\zeta$ set by the Planck scale (c.f. (\ref{eq:RedPlanck})). In fact, in order for $M_{Pl}$ not to exceed $10^{19}$ GeV, $\zeta$ should be $\lesssim 5.7$.

\subsection{Contribution to Non-Newtonian Gravity}
An important aspect of the RS model, is that gravity will be modified at short distances due to the exchange of KK gravitons in addition to the zero mode \cite{Randall:1999vf}. In addition, the exchange of the radion itself will also modify gravity. Therefore it is important to calculate these contributions in order to be consistent with fifth force searches. A rigorous approach for calculating the contributions to non-Newtonian gravity in the RSI model was presented in \cite{Callin:2004py}. Here we will summarize the main points for obtaining the corrections and relegate the detailed derivation to Appendix \ref{App:NonNewtonian}.

For any two particles, the gravitational force they feel is due to their exchange of the massless gravition, the massive KK graviton modes and the radion\footnote{I ignore the exchange of the radion's KK tower as its effect is subleading compared to the radion itself.}. Therefore, the gravitational corrections can be extracted from the propagators of the following processes at vanishing energy transfer:

\begin{equation}\label{eq:propagators} 
      U(r) = \displaystyle{\lim_{k^{0} \to 0}} 
            \qquad \includegraphics[width=0.6\linewidth, valign=c,]{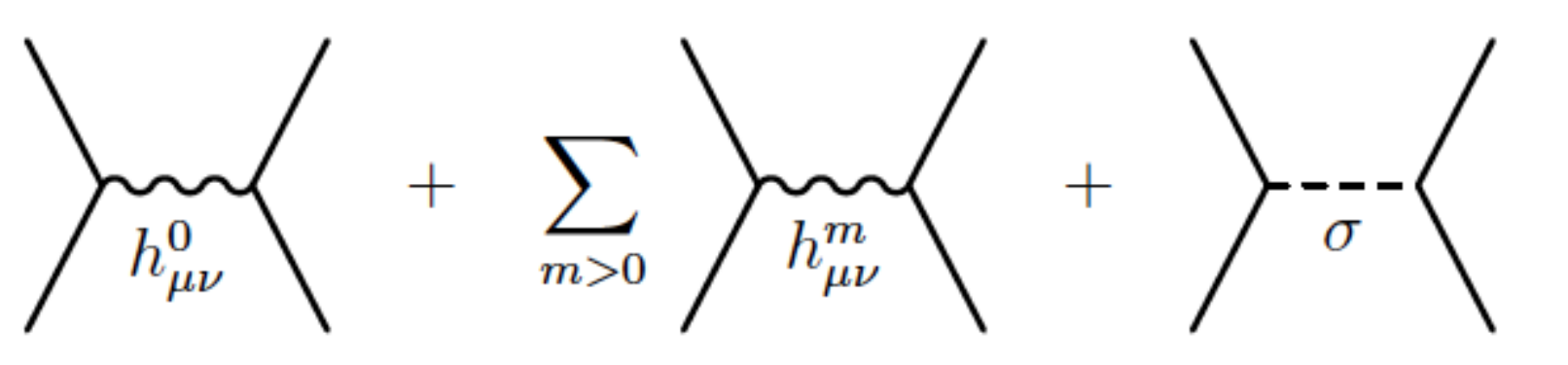} 
\end{equation}        

The first term gives the usual Newtonian potential, while second and third are Yukawa-type potentials exponentially-suppressed by the masses of the KK gravitons and the mass of the radion respectively. To find the corrections to the Newtonian gravity, two particles of masses $M_{1,2}$ are assumed to be localized on the IR brane exchanging the gravitons and radion while at rest. The Feyman diagrams are calculated in the limit of vanishing energy transfer and the full gravitational potential is extracted (see  Appendix \ref{App:NonNewtonian} for details): 
\begin{equation}\label{eq:FullGravPotential}
U(r) = \frac{G_{N}M_{1}M_{2}}{r} \Bigg[1+\frac{4}{3}\frac{|\Psi_{1}(m_{1},y_{1})|^{2}}{e^{\zeta m_{0}r}-1} +(\frac{4\sqrt{2}}{\kappa \Lambda_{IR}})^{2}  e^{-m_{\sigma} r} \Bigg]
\end{equation}
The second term gives a negligible correction compared to the Newtonian potential due to the exponential suppression of the heavy masses of the KK gravitons, however, the radion's contribution cannot be neglected as it's mass can be very small and almost vanishing. In the original RS proposal, the radion is massless, leading to a long-range force which contradicts experiment. This was overcome by the GW mechanism, which gives mass to the radion thereby avoiding this dilemma. However, with the GW mechanism, the modification to gravity can be significant at short distances due to the strong coupling of the radion and KK gravitons which would be $\sim O(\TeV^{-1})$. In this model, although the radion can be very light, it's much lower coupling to matter prevents it from significantly modifying gravity. In other words, although the mass of the radion becomes lighter for smaller $\epsilon$, it's coupling to matter also decreases with $\epsilon$ (see next section), thus remaining consistent with the bounds set by fifth force searches. I will discuss these experimental bound in more detail in Section \ref{Chap:Bounds}, where it will be shown that for the mass range of interest, those bounds do not exclude the model.

\section{Radion Couplings, Decays and Lifetimes}\label{Chap:Lifetimes}
\subsection{Radion Couplings}\label{Sec:Couplings}
The coupling of the radion to matter depends on where this matter is localized. The radion couples to brane-localized matter through the trace of the stress-energy tensor \cite{Davoudiasl:1999jd}:
\begin{equation}\label{eq:RadionBraneCoupling}
\mathcal{L}_{brane} = \frac{\sigma(x)}{\Lambda_{i}} \Tr T_{\mu\nu}
\end{equation}
where $\Lambda_{UV/IR}$ are the radion's UV and IR couplings respectively. On the other hand, the coupling to matter propagating in the bulk is determined by the overlap integral of the wavefunctions' profiles in the extra dimension. Throughout this paper, I will assume that the Higgs boson and the top quark are localized on the IR brane, and that the rest of the fermions are localized on the UV brane, while allow the massless and massive gauge bosons to propagate in the bulk.

The coupling to matter localized on the UV brane is highly suppressed compared to the coupling to either IR matter or bulk fields. Therefore all fields on the UV branes can be safely neglected. In general, $\Lambda_{i}$ will have two contributions, one from the gravity sector and the other from the bulk scalar. The full contibution is given by (see \cite{Csaki:2000zn} for a complete treatment):
\begin{equation}\label{eq:DecayConstant}
\Lambda_{i} = \frac{\sqrt{6}}{\kappa} \frac{1}{F(y_{i})} \Big[ \int_{y_{0}}^{y_{1}} e^{-2A(y)} \Big( 2F^{2} + \frac{3}{\kappa^{2} \varphi^{'2}}(F^{2}-2A' F)^{2} \Big) \Big]^{1/2}
\end{equation}
where $y_{0,1}$ gives the coupling to the UV and IR branes respectively. The dominant part of the radion's coupling to massive gauge bosons is given by\cite{Csaki:2007ns}\footnote{Here the contribution is doubled due to orbifolding.}:
\begin{equation}\label{eq:MassiveCoupling}
\mathcal{L}_{massive} = -\frac{\sigma(x)}{\Lambda_{i}} \Big[ 4M_{W}^{2} W_{\mu}^{+} W^{\mu -} +2 M_{Z}^{2} Z_{\mu}Z^{\mu}  \Big]
\end{equation}

For massless gauge bosons in the bulk, there are additional contributions to the coupling coming from the trace anomaly and triangle diagrams. The coupling to photons and gluons is given by \cite{Csaki:2007ns, Bellazzini:2012vz, Abu-Ajamieh:2017khi, Chacko:2012sy, Chacko:2014pqa}:
\begin{equation}\label{eq:MasslessCoupling}
\mathcal{L}_{massless} = \frac{\alpha}{8 \pi} \Big(b_{IR}^{EM} - b_{CFT}^{EM} \Big) \frac{\sigma(x)}{\Lambda_{IR}} F_{\mu\nu}F^{\mu\nu} + \frac{\alpha_{s}}{8 \pi} \Big(b_{IR}^{(3)} - b_{CFT}^{(3)} \Big) \frac{\sigma(x)}{\Lambda_{IR}} G_{\mu\nu}^{a}G^{a\mu\nu}
\end{equation}
where $b^{EM}$, $b^{(3)}$ are the $\beta$-function coefficients of the SM $U(1)$ and $SU(3)$ respectively. For this particular choice of fields localization, $b^{EM} = -\frac{2}{3}$ and $b^{(3)} = -\frac{19}{9}$. Here, $b_{CFT}$ is the CFT contribution to the running and is given by:
\begin{equation}\label{eq:bCFT}
b_{CFT} = -\frac{16 \pi^{2}R}{g_{5}^{2}}
\end{equation}
where $g_{5} = \sqrt{2(y_{1}-y_{0})}g_{4}$ is the 5D gauge coupling. Notice that the radion's coupling to the $W$ and $Z$ receives contributions from (\ref{eq:MasslessCoupling}) through their kinetic term in addition to (\ref{eq:MassiveCoupling}), however, the latter is the dominant contribution, see \cite{Bellazzini:2012vz} for more details. Since the radion's decay to massive gauge bosons is kinematically forbidden, I will ignore them and focus only on the latter two. Notice also that (\ref{eq:MasslessCoupling}) provides the radion's coupling to massless gauge bosons in any general RS model scenario whether it implements the CPR mechanism or not. To be more precise, eq. (\ref{eq:MasslessCoupling}) gives the dilaton's coupling to massless gauge bosons propagating in the bulk.

The final piece we need for our calculation is the radion's coupling to nucleons and pions. If the radion is heavy enough, it could decay to a pair of nucleons or pions. In addition, the coupling to nucleons could affect the cosmological bounds on the radion (see Section \ref{Chap:Bounds}). The radion couples to nucleons \cite{Shifman:1978zn,Burgess:2000yq,Andreas:2008xy, Kribs:2009fy,Cheng:2012qr} and pions \cite{Voloshin:1985tc, Barbieri:1988ct} through quarks and gluons similar to the Higgs. This calculation was adapted for the radion in \cite{Abu-Ajamieh:2017khi}:
\begin{equation}\label{eq:RadionNucleon}
\mathcal{L}_{\sigma NN} = \frac{b_{CFT}^{(3)}-b_{IR}^{(3)}}{b_{UV}^{(3)}} \frac{m_{N}}{\Lambda_{IR}} \sigma(x) \overline{N}N \equiv g_{\sigma NN} m_{N} \sigma(x) \overline{N}N
\end{equation}
where $m_{N}$ is the nucleon mass and $b_{UV}^{(3)}$\footnote{In \cite{Abu-Ajamieh:2017khi} $b_{UV}^{(3)}$ is called $b_{light}^{(3)}$.} is the $\beta$-function coefficient of quarks localized on the UV brane and is equal to $-\frac{10}{3}$ in this model. Note that $g_{\sigma gg}$, $g_{\sigma \gamma \gamma}$ and $g_{\sigma NN}$ all have dimension $(mass)^{-1}$.

\begin{figure}[!ht]
 \centering
  \includegraphics[height=.3\textheight]{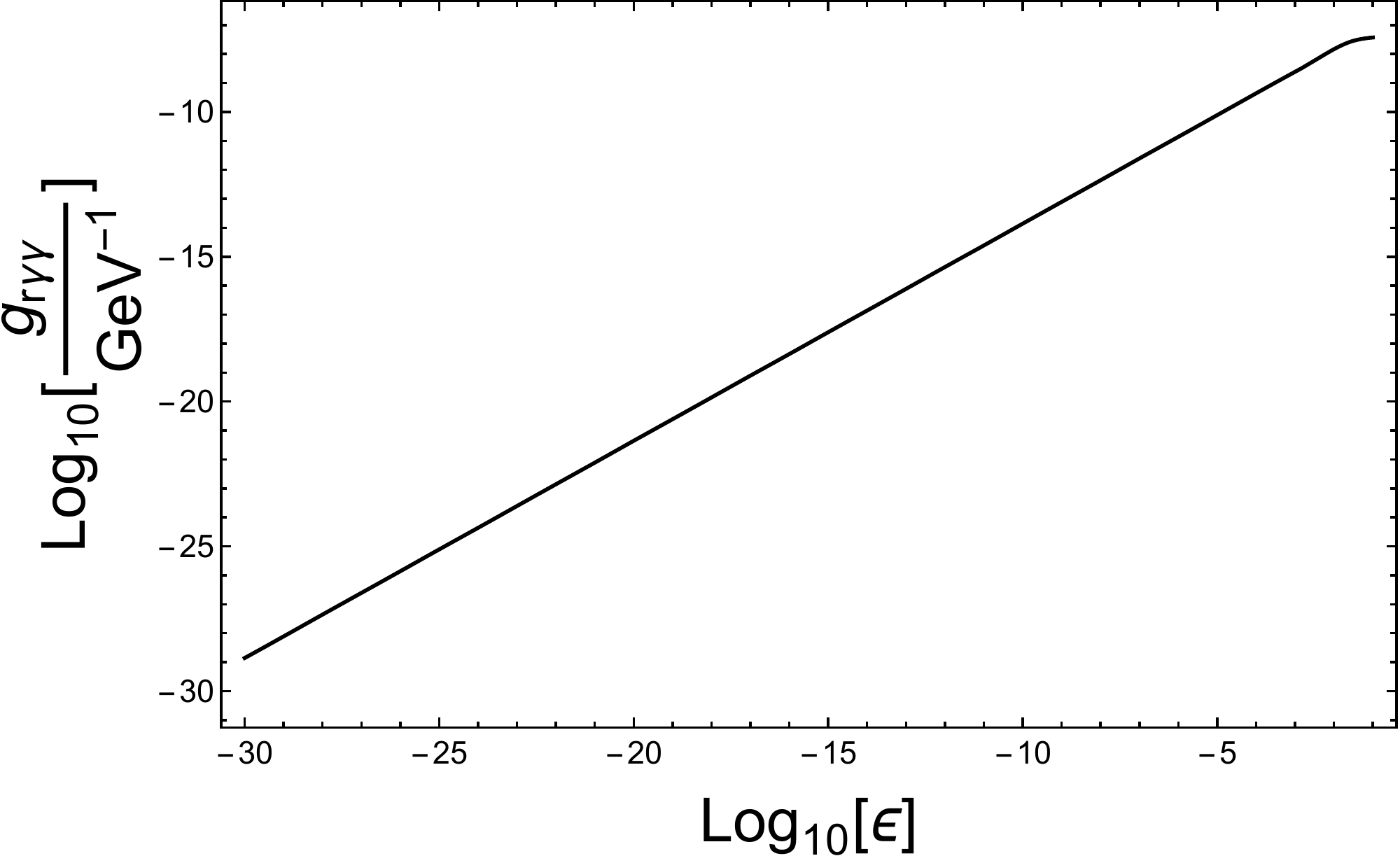}
  \caption{$\log_{10}{g_{\sigma\gamma\gamma}}$ Vs. $\log_{10}{\epsilon}$ for $\zeta =1.6$. The relationship is linear for small $\epsilon$.}
  \label{fig:gVsEpsilon}
\end{figure}

An important aspect of this model is that the coupling of the radion to SM matter becomes smaller for small $\epsilon$ (i.e. for small radion masses), and can be lowered to $\ll \TeV^{-1}$. To see this explicitly, we use (\ref{eq:Solutions1}) and (\ref{eq:Solutions2}) to find a solution for $F(y)$ using eq. (\ref{eq:RadionEOM}) subject to the boundary condition (\ref{eq:RadionBCs}), then we use it to perform the integral (\ref{eq:DecayConstant}) to find the decay constant $\Lambda_{IR}$. An analytical solution of $F(y)$ is not possible, so instead we do the calculation numerically. The relationship between the radion's coupling to photons and $\epsilon$ is shown in Fig. (\ref{fig:gVsEpsilon}). The plot shows a linear relationship between the $\log{g_{\sigma \gamma \gamma}}$ and $\log{\epsilon}$ for small $\epsilon$, which allows for the suppression of the coupling through making the running slower. A suppressed coupling to matter is crucial in order to make the radion stable without the need to resort to any parity argument to prevent its decay. This fact comes in contrast to the original RS model where the radion couples to IR matter with strength $\sim O(\TeV^{-1})$ irrespective of its mass. The reason for this is that with a smaller running of the scale-breaking operator, the radion's wavefunction will grow more slowly along the extra dimension, thereby making it less peaked near the IR brane and more "spread-out" across the bulk. Thus there will be less overlap between the radion's wavefunction and the IR brane, which reduces the radion's coupling to IR fields. The same effect happens for bulk fields. Another way to look at this is as follows:  in order to keep the size of the extra dimension that corresponds to the hierarchy between the Planck scale and EW scale fixed for smaller $\epsilon$, we need to increase the (attractive) UV brane potential. This attractive UV potential acting on the radion will make its wavefunction less peaked in the IR region and more spread-out along the bulk.\footnote{Another possibility is to keep the UV potential fixed, and reduce the IR potential. which means less attractive force from the IR brane on the radion, which leads to the same effect, however, here I adopt the first tuning.}. We investigate the dependence of the radion's coupling on $\epsilon$ in more detail in Appendix \ref{App:Fdep}.

We should note however, that for our choice of field localization (where only the top quark and the higgs boson are localized on the IR brane), the coupling to IR matter will not have much impact of the phenomenology of the light radion in this paper, since as we show below, the radion decay to $\bar{t}t$ and $hh$ is kinematically forbidden. However, it is can be relevant for heavier radion masses if they can be made stable enough.

\begin{figure}[!ht]
 \centering
  \includegraphics[height=.3\textheight]{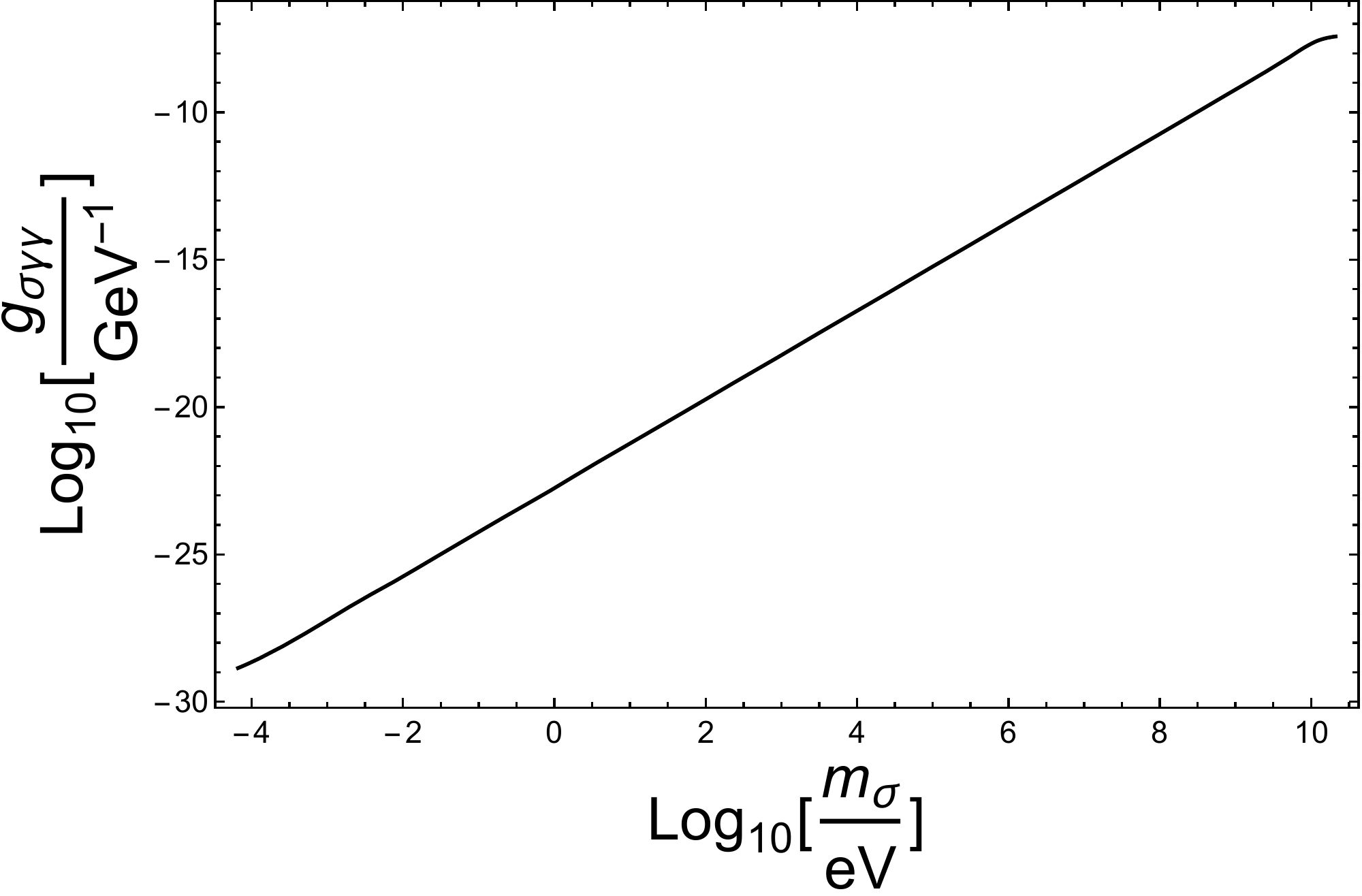}
  \caption{$\log_{10}{g_{\sigma\gamma\gamma}}$ Vs. $\log_{10}{m_\sigma}$ for $\zeta =1.6$. The relationship is linear for small masses.}
  \label{fig:gVsm}
\end{figure}

The radion's coupling to photons against it's mass is plotted in Fig. \ref{fig:gVsm}. The plot shows a linear relationship between the $\log{g_{\sigma \gamma \gamma}}$ and $\log{m_{\sigma}}$ for small $\epsilon$. So in essence, the radion's mass can be made very light, and its coupling to matter can be significantly suppressed. In introducing an external operator that breaks scale invariance through its running, one can avoid the dangerous cosmological consequences of an almost massless radion as in the orignal RS model. This can make the radion stable and potentially a good candidate for dark matter as we demonstrate below.

\subsection{Radion Decay and Lifetime}
Armed with the couplings, we can proceed to calculate the radion's decay widths and lifetime. The relevant decay widths are:
\begin{equation}\label{eq:TopDecay}
\Gamma(\sigma \rightarrow \bar{t}t) = \frac{N_{c}}{8\pi} \frac{m_{\sigma} m_{t}^{2}}{\Lambda_{IR}^{2}} \Big[1-\frac{4 m_{t}^{2}}{m_{\sigma}^{2}} \Big]^{3/2}
\end{equation}
\begin{equation}\label{eq:HiggsDecay}
\Gamma(\sigma \rightarrow hh) = \frac{9}{8 \pi} \frac{m_{h}^{4}}{m_{\sigma}\Lambda_{IR}^{2}} \Big[1-\frac{4m_{h}^{2}}{m_{\sigma}^{2}} \Big]^{1/2}
\end{equation}
\begin{equation}\label{eq:MasslessDecay}
\Gamma(\sigma \rightarrow gg/ \gamma\gamma) =N_{c} \frac{g_{gg/ \gamma\gamma}^{2}}{4\pi} m_{\sigma}^{3}
\end{equation}
\begin{equation}\label{eq:NucleonDecay}
\Gamma(\sigma \rightarrow \overline{N}N) = \frac{g_{\sigma NN}^{2}}{8\pi} m_{\sigma} m_{N}^{2} \Big[1-\frac{4m_{N}^{2}}{m_{\sigma}^{2}} \Big]^{3/2}
\end{equation}
\begin{equation}\label{eq:PionDecay}
\Gamma(\sigma \rightarrow \pi^{+}\pi^{-}) = \frac{g_{\sigma NN}^{2}}{16\pi} m_{\sigma}^{3} \Big(1-\frac{4m_{\pi}^{2}}{m_{\sigma}^{2}} \Big)^{\frac{1}{2}} \Big(1+\frac{m_{\pi}^{2}}{m_{\sigma}^{2}} \Big)^{2}
\end{equation}
where $N_{c}$ is a color factor. In order for the radion to be a good dark matter candidate, it's lifetime should be larger than or at least comparable to the age of the universe. Using the decay formulas above, we can find the lifetime of the radion for the range of $\epsilon \in [10^{-30},10^{-1}]$, which corresponds to a mass range $\sim 10^{-5} \eV - 10 \GeV$.

\begin{figure}[!ht]
 \centering
  \includegraphics[height=.3\textheight]{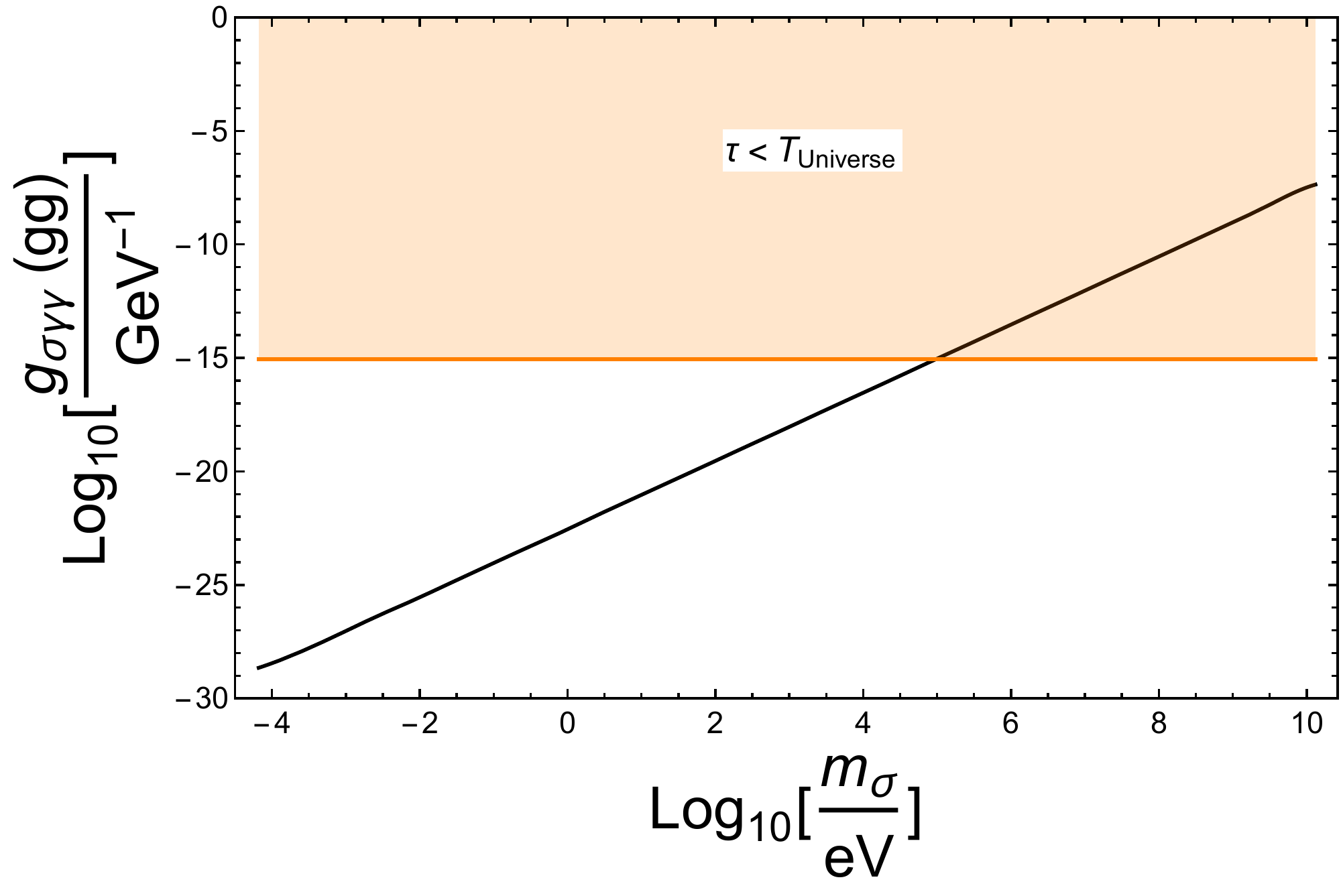}
  \caption{Allowed region in the $g_{\sigma \gamma\gamma} -m_{\sigma}$ parameter space. The excluded region corresponds to a radion lifetime less than the age of the universe.}
  \label{fig:gVsmUni}
\end{figure}

For this mass range, decays to $\bar{t}t$ and $hh$ are not kinematically allowed. For radion masses $\sim O(\GeV)$, decays to pions and nucleons dominate. On the other hand, for $m_{\sigma} < 2m_{\pi}$, only decays to photons can occur. For radion masses $\sim O(\MeV) - O(\GeV)$, the radion decays very quickly and that region of parameter space is excluded. Only for lighter masses can the the lifetime be long enough. Therefore, the decay to photons is the only relevant process. Fig. \ref{fig:gVsmUni} shows the allowed region in the $g_{\sigma \gamma\gamma} - m_{\sigma}$ parameter space. The shaded area corresponds to decays faster than the age of the universe and therefore that region is excluded. As can be seen from the figure, the radion can serve as a DM particle if its mass  $\lesssim 100 \keV$ and its coupling  $ \lesssim 10^{-15} \GeV^{-1}$, which corresponds to $\epsilon \lesssim 10^{-12}$. For this mass range, the radion behaves like an Axion-Like Particle (ALP) rather than a Weakly-Interacting Massive Particle (WIMP). This again is expected, since both the coupling and the mass of the radion are proportional to $\epsilon$, which means that the range of suppressed couplings that would make the radion stable would necessarily correspond to light masses.

\section{Experimental Bounds}\label{Chap:Bounds}
In this section, I will discuss the relevant experimental and cosmological bounds on DM radion. As was shown previously in Fig. \ref{fig:gVsmUni}, for the radion to be a good DM candidate, it has to have a mass $\lesssim 100 \keV$ and a coupling to SM matter $\lesssim 10^{-15} \GeV^{-1}$. In this region of parameter space, the relevant experimental bounds come from cosmological ALP bounds and non-Newtonian fifth force experiments.

Using the results from \cite{Cadamuro:2011fd}, we find that the relevant cosmological constraints come from the WMAP measurement of DM fraction, Extragalactic Background Light (EBL) and X-Rays from galactic spectra. The constraints from WMAP stem from the argument that DM ALPs should not exceed the DM fraction measured by WMAP. EBL bounds come from the fact that the photons produced by ALP decays when the universe is transparent must not exceed the extragalactic background light, while the argument behind X-Rays bounds is that photons from ALPs decaying in galaxies should not exceed the background in the galactic spectra. 

The most stringent bounds come from X-Rays observations, and they constrain the mass of the radion to be $\lesssim 1$ KeV with coupling $\lesssim 10^{-16}$ GeV$^{-1}$. In this model, a radion mass of $\sim 1$ KeV corresponds to $\epsilon \approx10^{-15}$ and has a coupling $\sim 10^{-18}$ GeV$^{-1}$.  We should also note that there is a more stringent bound on Hot Dark Matter (HDM) produced via freeze-out coming from the WMAP observation of the the DM fraction. If the DM is a hot relic produced via freeze-out, then the observed relic abundance sets an upper limit on the mass of $\sim 160$ eV, corresponding to $\epsilon \sim 10^{-17}$. This limit comes from the fact that when a DM particle freezes-out while still relativistic (i.e. when $x_{f} = \frac{m}{T_{fo}} \lesssim 3$), then their relic abundance will be given by \cite{Kolb}:
\begin{equation}\label{HDMRelicAbundance}
\Omega h^{2} = 7.83 \times 10^{-2} \frac{g_{eff}}{g_{*s}} (m/ \eV)
\end{equation}
where $g_{eff}$ is the effective number of degrees of freedom ($=1$ for a scalar) and $g_{*s}$ measures the total number of effectively massless degrees of freedom at the freeze-out temperature $T_{fo}$. One can see that for a measured value of $\Omega_{DM}h^{2} =0.12$ that an upper bound of $\sim 160$ eV on the mass is obtained. This however, only applies to HDM produced via freeze-out. In this paper, we will assume that the radion is not produced via freeze-out, but instead is produced via freeze-in. Therefore this bound will not apply and we can set the upper bound on the mass of the radion from cosmological observation to be $\sim 1$ KeV.

The other source of bounds on the allowed masses and couplings of the radion comes from fifth force searches. As discussed in Section \ref{Chap:Gravity}, the radion and the massive KK gravitons will have additional contributions to the Non-Newtonian potential. Such contributions should not exceed the limits set by fifth force searches. We can use eq. (\ref{eq:FullGravPotential}) to put the non-Newtonian potential in the form
\begin{equation}\label{eq:NonNewtonianForm}
U(r) = \frac{G_{N}M_{1}M_{2}}{r} \Big(1+ |\alpha| e^{-r / \lambda}  \Big)
\end{equation}

\begin{figure}[!ht]
 \centering
  \includegraphics[width=.47\textwidth]{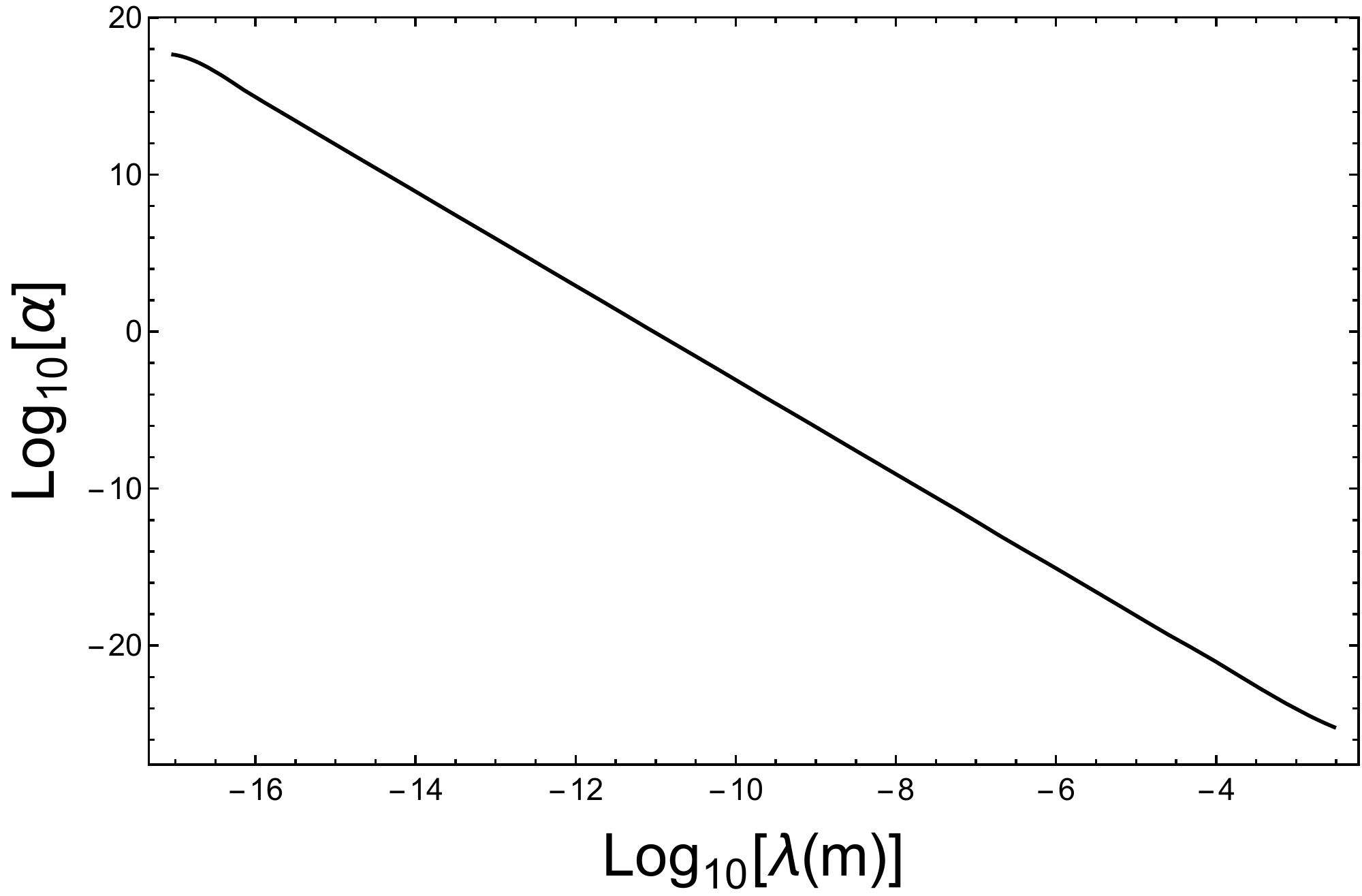}
   \includegraphics[width=.45\textwidth]{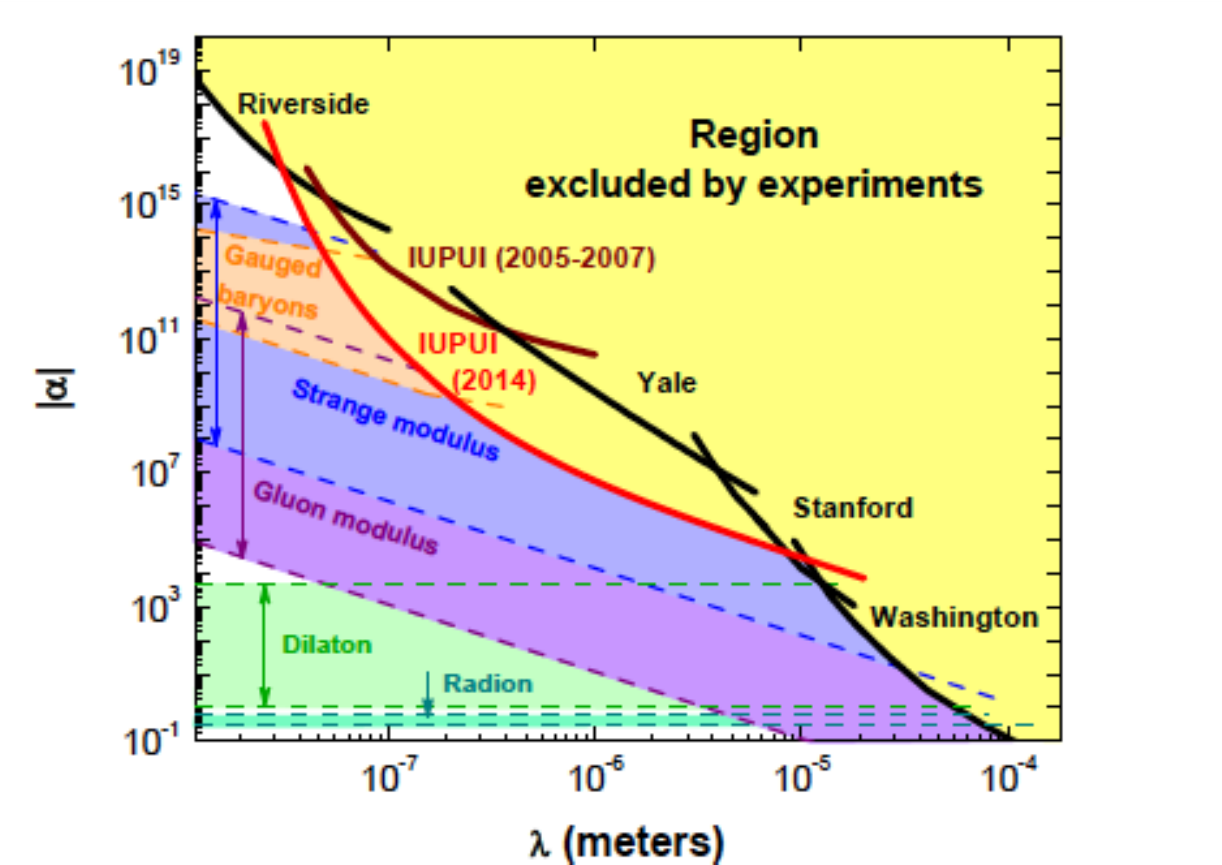}
  \caption{(Left): The radion's contribution to non-Newtonian gravity for $\zeta = 1.6$, (Right): The latest bounds from non-Newtonian gravity searches conducted by Y. J. Chen et. al.. This plot also shows the regions in parameter space relevant to existing radion and dilaton models.}
  \label{fig:NonNewtonianGrav}
\end{figure}

We can use this in order to compare with the latest results of fifth force experiments. The left panel of Fig. \ref{fig:NonNewtonianGrav} shows  the radion contribution to non-Newtonian gravity neglecting the minuscule effect of the KK gravitons, while the right panel shows the latest experimental bounds found in \cite{Chen:2014oda}. Comparing between the two plots, we see that this model easily escapes these bounds. In addition, the right panel shows the region in parameters space relevant to existing radion and dilaton models (the light and dark green regions). As can be seen from the plot, current models predict a radion/dilaton coupling of $\sim 10^{-1} - 10^{3}$ realtive to gravity, whereas in this model the relative coupling is predicted to be much smaller ($\sim 10^{-10} - 10^{-20}$ for distances between $\sim 10$ nm and $0.1$ mm). This is largely due to the significantly lower radion couplings which stems from the slow running. In view of this, it is unlikely that similar searches in the near future would constrain this model.

\section{Relic Abundance}\label{Chap:RelicAbundance}
\begin{figure}[!ht]
 \centering
  \includegraphics[width=0.7\textwidth]{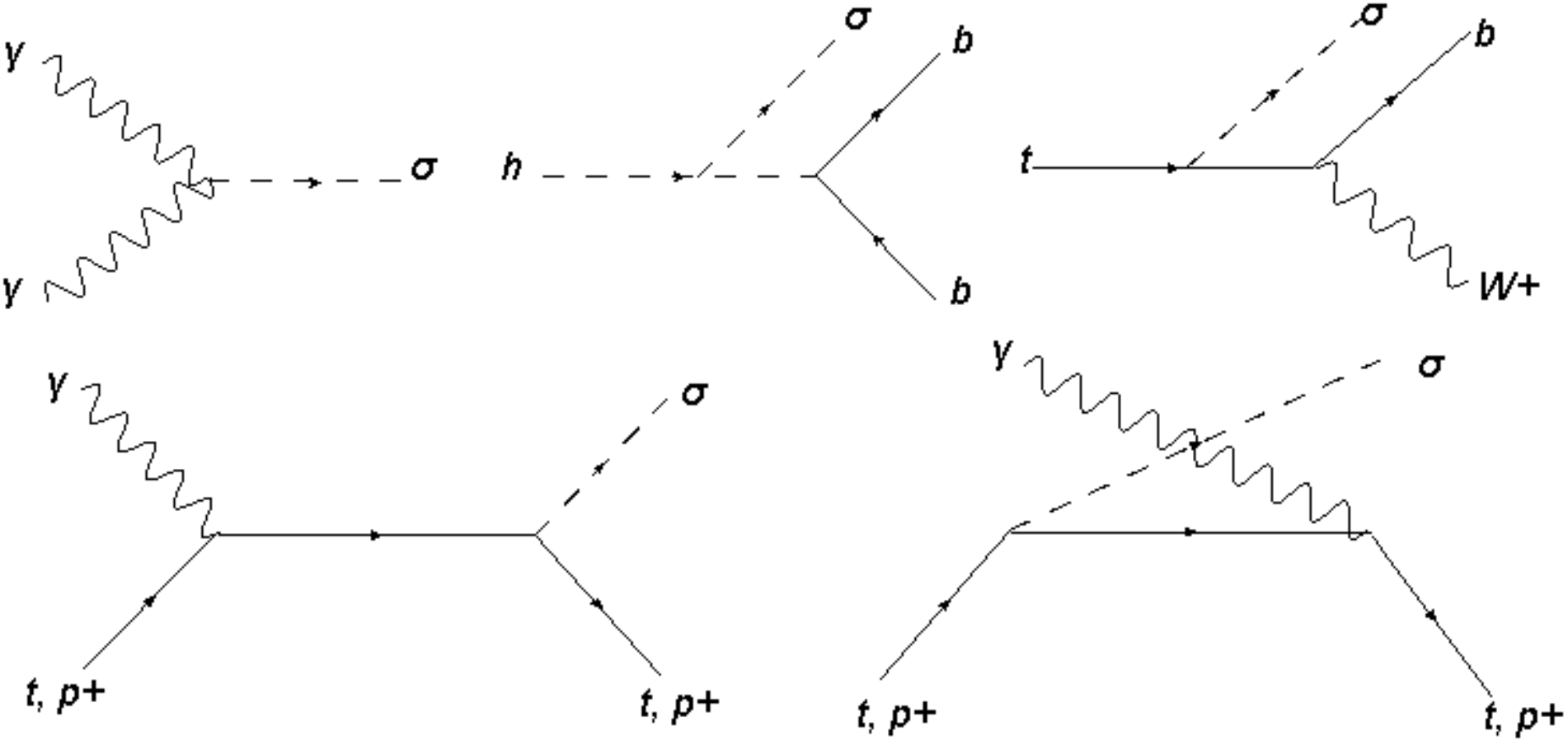}
  \caption{The main radion production processes.}
  \label{fig:ProductionProcesses}
\end{figure}

In this section, I will discuss the possible processes for producing the readion and try to calculate the corresponding relic abundance. Given that only the top and the higgs are localized on the IR brane, and that massless gauge bosons are allowed to propagate in the bulk, the dominant radion production processes are summarized in Fig. (\ref{fig:ProductionProcesses}). As it turns out, for the mass range or interest, the radion will always be decoupled, i.e. its interaction rate $\Gamma$ will always be less than the rate of the expansion of the universe $H$. This is to be expected since the coupling in the range of interest is very small. Therefore the radion DM cannot be produced via freeze-out. However, the decoupling of the radion suggests that could be a Feebly-Interacting Massive Particle (FIMP) produced via freeze-in \cite{Hall:2009bx}.

At low temperatures $\sim \MeV$, $\gamma p \rightarrow \sigma p$ dominates, so I will focus on the relic abundance produced through this process. As shown below, most of the relic abundance is produced at temperatures below the proton mass, therefore, we can assume that the proton's initial and final momenta are negligible, such that $E_{p} \approx m_{p} \gg E_{\gamma}$. In this regime, the cross section simplifies to:

\begin{equation}\label{eq:XSection}
\sigma({\gamma p \rightarrow \sigma p}) \approx \frac{e^{2}g_{\sigma NN}^{2}}{4\pi s}
\end{equation}

\begin{figure}[!ht]
 \centering
  \includegraphics[width=.43\textwidth]{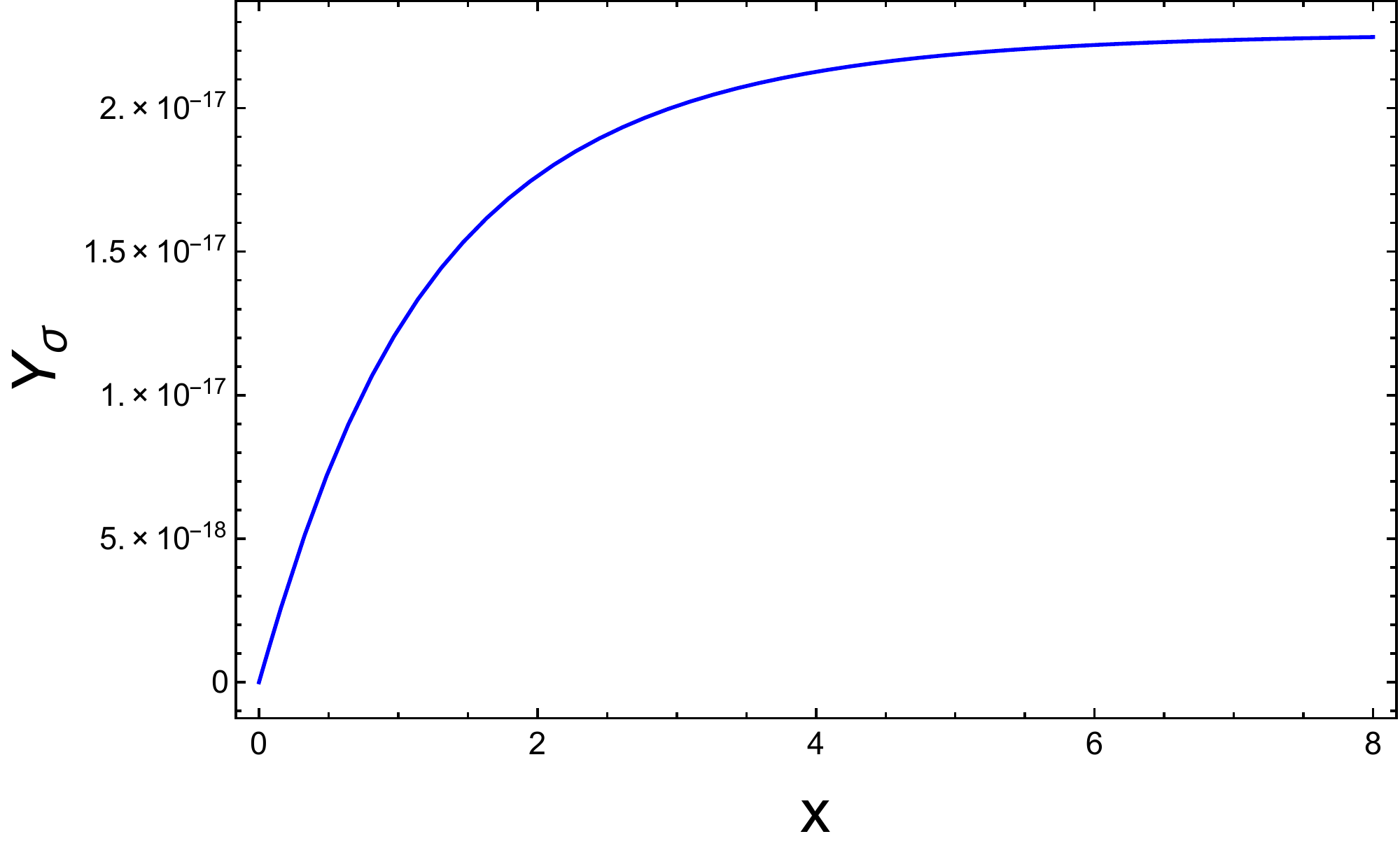}
  \includegraphics[width=.4\textwidth]{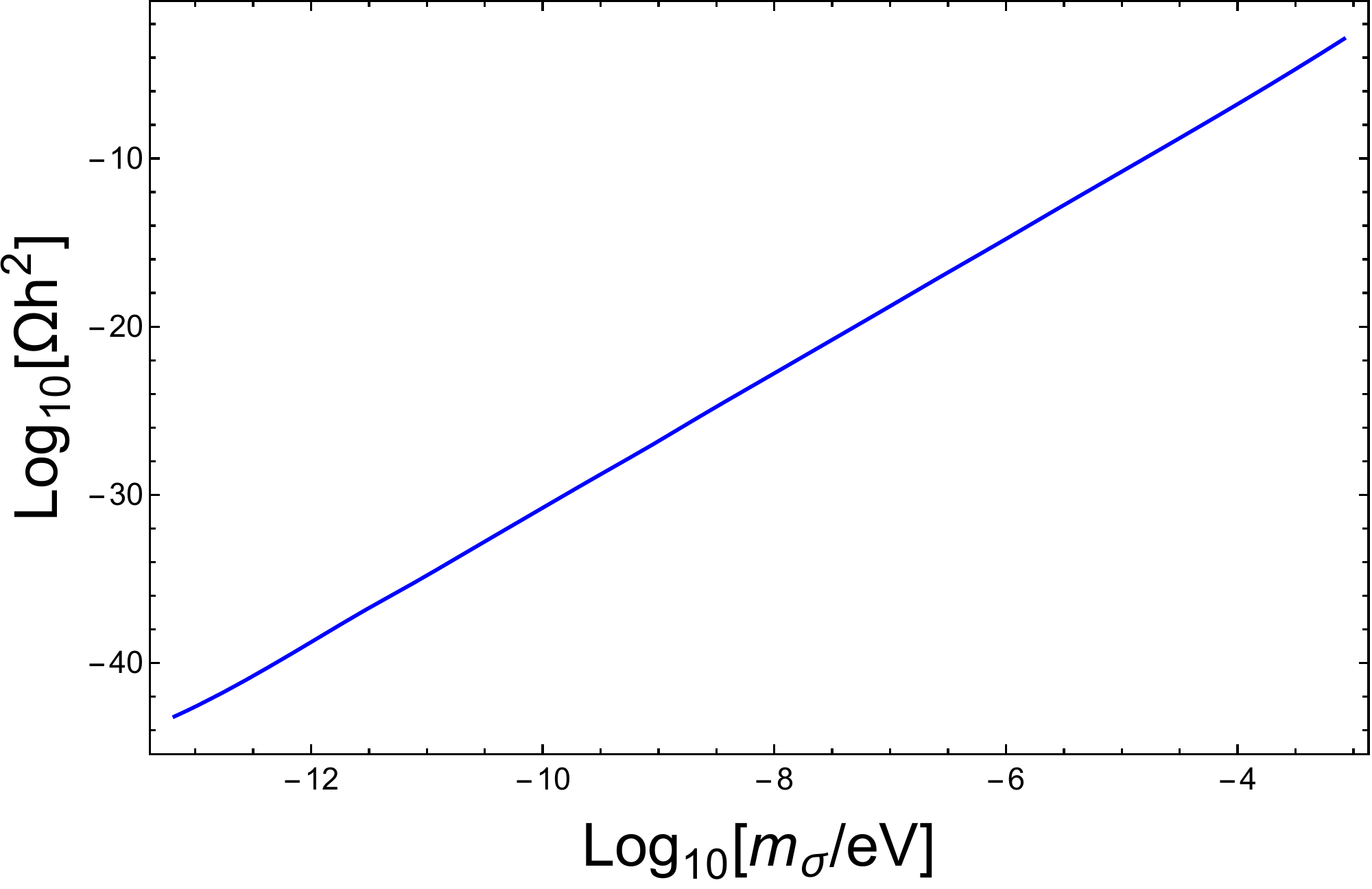}
  \caption{(Left): The yield as a function of $x = m_{p}/T$.  (Right): The relic abundance as a function of the radion mass. }
  \label{fig:YieldandRelic}
\end{figure}

Using this, we can find the yield and relic abundance (see Appendix \ref{App:RelicAbundance} for details):

\begin{dmath}\label{eq:Boltzmann3_0}
\frac{dY}{dx} \approx 1.7171 \times 10^{14}\frac{g_{\sigma NN}^{2}}{g_{*s}\sqrt{g_{*}}} \frac{x}{K_{2}(x)}  \MeijerG{3,}{0}{1,}{3}{1}{-\frac{3}{2},-\frac{1}{2},\frac{1}{2}}{\frac{x^{2}}{4}}
\end{dmath}

\begin{equation}\label{eq:FinalRelic_0}
\Omega h^{2} \approx 3.325 \times 10^{27} \frac{m_{\sigma}}{M_{Pl}} Y_{\infty}
\end{equation}
where $x=m_{p}/T$, $\MeijerG{m}{n}{p}{q}{a_1,\ldots,a_p}{b_1,\ldots,b_q}{z}$ is the Meijer function and $Y_{\infty}$ is the final yield. We can use (\ref{eq:Boltzmann3_0}) to plot the yield as a function of $x$. The left panel of Fig. (\ref{fig:YieldandRelic}) shows that most of the radion relic is produced for $x \lesssim 4$, which corresponds to a temperature of $\sim 230$ MeV. This makes the radion in the mass range $\lesssim 1$ KeV, a HDM candidate. The right panel shows the radion relic abundance fraction. Given that the observed relic abundance is $\Omega h^{2} =0.12$, and that the fraction of HDM should not exceed $35 \%$ \cite{Boyarsky:2008xj}\footnote{HDM cannot make up the entire DM in the uiverse since that would be inconsistent the observed structure formation. The above fraction is obtained from the latest Lyman-$\alpha$ forest data. See \cite{Boyarsky:2008xj} for more details.}, we can see from the plot that the relic abundance produced this way is too small to account for all the HDM content of the universe. Including the other processes in Fig. (\ref{fig:ProductionProcesses}) will not be sufficient to increase the overall relic abundance significantly. Therefore, if we insist on accounting for a larger portion of the DM fraction, alternative production mechanisms need to be considered.

One possible way to produce a larger fraction is to relax the assumption that the radion's initial abundance is negligible, i.e.  that a significant number of radions is produced early in the universe. In this case, its decoupling would guarantee that the initial abundance will not annihilate significantly, and that it will mostly survive to the current epoch. We can achieve this via proper UV completion of the theory. For instance, if we introduced heavy field strongly coupled to the radion, more abundance would be produced at high energies before it quickly decouples leading to a larger relic fraction. Another possibility that not only could produce a larger relic, but also could lead the radion being CDM is to assume that the radion forms a Bose-Einstein Condensate (BEC). Since the radion resides in an effective potential, it could form a BEC and therefore produce cold relics. However, forming a BEC might lead to the oscillation of the mass scales of the theory (e.g. the Planck mass), an issue that needs to be addressed if this possibility is to be viable. One possible way to avoid this is to make sure that the period of oscillation is long enough so as not to contradict any cosmological observations. I will postpone tackling these issues to a future work.

\section{Contribution to the Cosmological Constant}\label{Chap:CosmConst}
As highlighted in \cite{Bellazzini:2013fga}, an interesting aspect of SBSI via slow running is the suppressed contribution to the cosmological constant. As it is well-known,  the value of the effective potential at its minimum in any gravitational theory corresponds to its contribution to the cosmological constant. More specifically, we can split the cosmological constant into a UV component that corresponds to the UV cutoff scale, a TeV contribution near the IR brane corresponding to the contributions from the phase transition of a conformal theory into the broken phase, and an IR contribution that includes all contributions below the EW scale (such as the  QCD phase transition).

In general, the TeV contribution is $\sim O(\TeV^{4})$ and is usually the most problematic part. However, as we showed in (\ref{eq:VIRmin}), this contribution is suppressed by $\epsilon$. For $\epsilon \approx 4 \times 10^{-3}$, this contribution becomes comparable to the QCD phase transition $\sim O(\Lambda_{QCD}^{4})$. On the other hand, yielding a value comparable to the measured cosmological constant of $10^{-35} \MeV^{4}$ would require $\epsilon \lesssim 10^{-60}$, a value achievable in this model. Although such a value might look fine-tuned, we should keep in mind that we are actually tuning the UV brane potential against the IR potential. We can estimate the amount of tuning as $\frac{v_0 (\epsilon = 10^{-60})}{v_{1}} \approx  47\%$, which is not significant at all. The reason for this is that the minimum of the effective potential is insensitive to the running at small values of $\epsilon$ (see the left panel of Fig. \ref{fig:V0andF0}). That is to say, when the running is very slow, it will have no effect on setting the size of the extra dimension, which would be determined solely by the branes' tensions whose values are set by the hierarchy between the Planck scale and the EW scale. However, we should keep in mind that it remains a key assumption in this paper that in order for the radion to be light and stable enough to serve as a DM candidate (and for the contribution to the cosmological constant to be reduced), the scale breaking operator must be very close to marginality, i.e. we need to tune $\epsilon \ll 1$. For more discussion about the naturalness of models with small running, see \cite{Bellazzini:2012vz}.
	
\begin{figure}[!ht]
 \centering
  \includegraphics[height=.3\textheight]{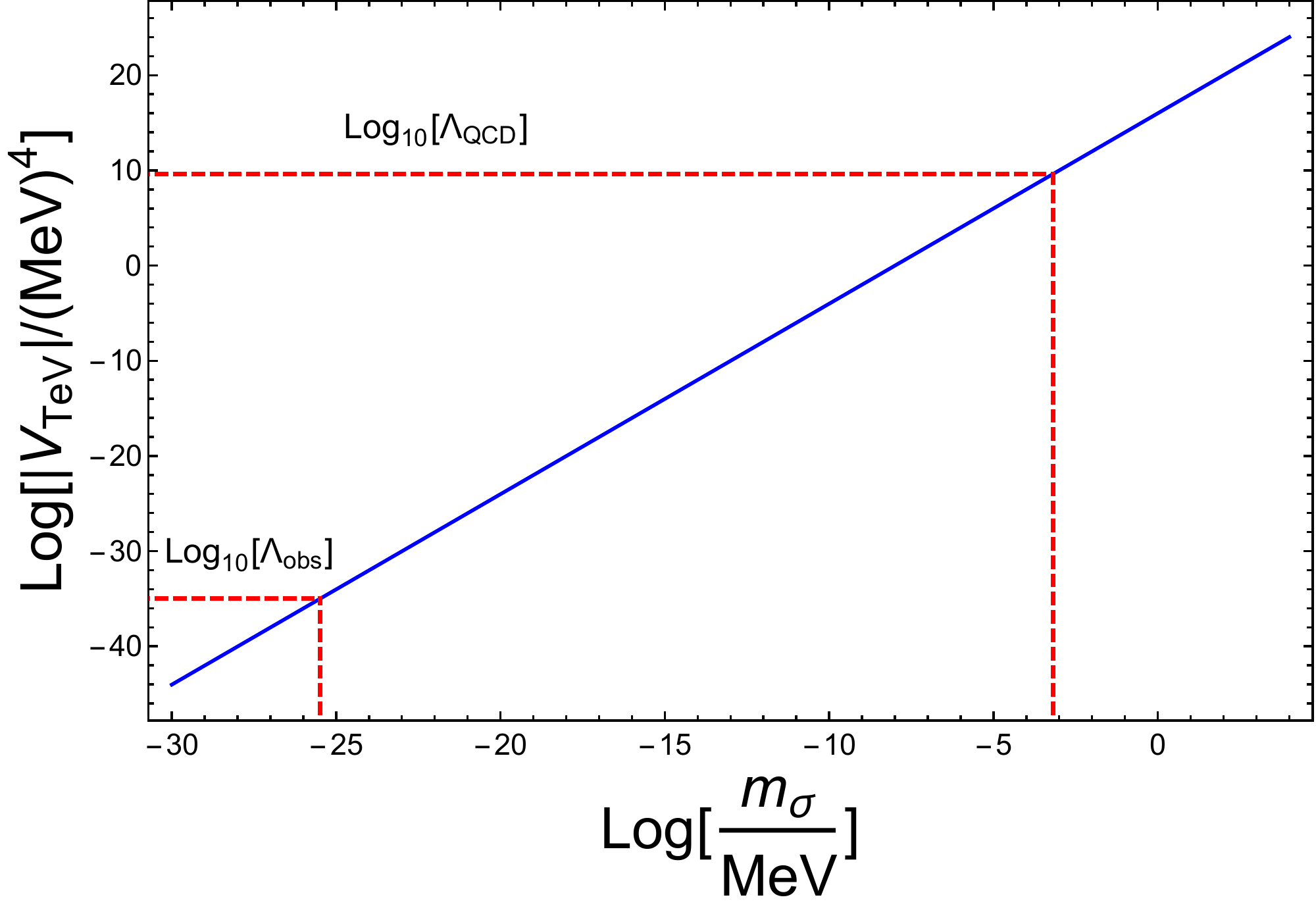}
  \caption{The IR contribution to the cosmological constant. The vertical lines show the radion masses that correspond to a contribution equal to the contribution from the QCD phase transition and to the observed value of the cosmological constant $ = 10^{-35} \MeV^{4}$.}
  \label{fig:CosmologicalConst}
\end{figure}

In Fig. \ref{fig:CosmologicalConst} we show the contribution to the cosmological constant corresponding to the radion mass. Note that in the region where the radion is a DM particle, the contribution $\ll \Lambda_{QCD}^{4}$. We can also see that a contribution comparable to the observed cosmological constant would require an radion mass of $\sim 10^{-25}$ eV. If the radion can form a BEC, then this small mass can pose the interesting possibility that the radion might be a Fuzzy Cold Dark Matter (FCDM) particle\cite{Hu:2000ke}, however, I will not pursue this possibility any further in this paper.

\section{Discusssion and Conclusions}\label{Chap:Conclusions}

The CPR mechanism can lead to a light dilaton/radion with suppressed couplings to SM matter through the slow running of the scale-breaking operator, which would make a radion with mass $\lesssim 1$ KeV stable enough to be a DM candidate. The possibility of such a light radion is consistent with experimental observations from LHC searches, cosmology and fifth force searches. This would open new venues for radion  phenomenology and extend the viability of the RS model beyond the current searches, helping to explain the absence of any signal corroborating the RS model thus far.

As the radion's contribution to non-Newtonian gravity is expected to be very small, fifth force search are unlikely to be the best means for experimentally investigating such models. On the other hand, LHC searches are an excellent means for constraining these models via searching for the associated KK gravitons and other KK modes. ALP searches, especially from cosmology, can help probe the relevant region of the parameters space as well.

In this paper we demonstrated that the slow running can reduce the contribution of the SBSI phase transition to the cosmological constant down to the observed value without fine-tuning of the branes' potentials. It was also shown that in this particular model that the radion could serve as a Hot Dark Matter FIMP produced via freeze-in mechanism due to its small mass and suppressed couplings. However, the resulting relic abundance of such a light radion is small to constitute a significant portion of the observed DM. One possibility to obtain a larger abundance is to assume that enough abundance is produced at high energies early in the universe. The decoupling of the radion would ensure that this initial abundance is not annihilated and that it survives today. Such a proposal would require proper UV completion of the model in order to non-thermally produce the required initial abundance. Another possibility is for the radion to form a BEC, rendering it a CDM candidate. For this possibility to be viable, one has to ensure that the oscillation of the scales does not contradict experiment, perhaps through making the period of oscillation long enough to be consistent with observation. There remains an open question of whether it would be possible to suppress the radion's couplings to matter at masses of $O(\GeV)$ without imposing any parity arguments, which would make the radion possibly a WIMP. This scenario runs into the difficulty that both the radion's mass and couplings are proportional to $\epsilon$, and will be severely constrained by the decay to the massless gauge fields in the bulk. Therefore, any solution would require special assumptions to make the radion's couplings $\ll \epsilon$. I will postpone tackling these issues to a future work.

\appendix
\section*{Acknowledgments}
I would like to thank my adviser John Terning for supervising this work. I also thank Markus Luty, Nemanja Kaloper, Lloyd Knox, Andreas Albrecht and Rui Zheng for the valuable discussions. I thank Emilija Pantic and Tessa Johnson  for answering my questions. I would also like to thank Y.-J. Chen, W. K. Tham, D. E. Krause, D. Lopez, E. Fischbach and R. S. Decca for granting permission to use their results.

\section{Contribution to Non-Newtonian Gravity}\label{App:NonNewtonian}
I will apply the treatment in \cite{Callin:2004py} in order to find the additional contribution of the radion and KK gravitons to the Newtonian potential. For any two massive particles, the gravitational contribution is given by the propagator of the gravitons and radion at vanishing energy transfer (c.f. (\ref{eq:propagators})). Focusing first on calculating the gravitational potential due to the KK gravitons including the zero mode; the 5D propagator is given by:
\begin{equation}\label{eq:5Dpropagator}
D_{\mu\nu\alpha\beta}^{(5)}(x,y;x'y') \equiv \Braket{0|T h_{\mu\nu}(x,y)h_{\alpha\beta}(x',y')|0} = \sum_{m} \Psi_{m}(y) \Psi_{m}(y') D_{\mu\nu\alpha\beta}^{(4,m)}(x,x')
\end{equation}
where $T$ indicates time-ordering and $D_{\mu\nu\alpha\beta}^{(4,m)}(x,x')$ is the 4D graviton propagator and is given by:
\begin{equation}\label{eq:4Dpropagator}
D_{\mu\nu\alpha\beta}^{4,m}(x,x') =\int \frac{d^{4}k}{(2\pi)^{4}} \frac{P_{\mu\nu\alpha\beta}^{(m)}(\vec{k})}{k^{2}-m^{2} +i \epsilon} e^{-i k(x-x')}
\end{equation}
where $P_{\mu\nu\alpha\beta}^{(m)}$ is the graviton's polarization tensor given by:
\begin{equation}
       P_{\mu\nu\alpha\beta}^{(m)}=
        \left\{ \begin{array}{ll}
            \frac{1}{2} \big(\eta_{\mu\alpha}\eta_{\nu\beta} +\eta_{\mu\beta}\eta_{\nu\alpha} - \eta_{\mu\nu}\eta_{\alpha\beta} \big) & (m=0) \\
         \\
         \frac{1}{2} \big(\eta_{\mu\alpha}\eta_{\nu\beta} +\eta_{\mu\beta}\eta_{\nu\alpha} - \eta_{\mu\nu}\eta_{\alpha\beta} \big) -\frac{1}{2m^{2}} \big(\eta_{\mu\alpha}k_{\nu}k_{\beta} +\eta_{\mu\beta} k_{\nu}k_{\alpha} \\ +\eta_{\nu\alpha} k_{\mu}k_{\beta} +\eta_{\nu\beta} k_{\mu} k_{\alpha} \big) + \frac{1}{6} \big(\eta_{\mu\nu} +\frac{2}{m^{2}} k_{\mu}k_{\nu} \big) \big(\eta_{\alpha\beta} +\frac{2}{m^{2}} k_{\alpha}k_{\beta} \big) & (m>0)
        \end{array} \right.
\end{equation}

For two particles localized on the IR brane, and using (\ref{eq:GravitonCoupling}), we find that the first two diagrams in  (\ref{eq:propagators}) yield:
\begin{equation}\label{eq:Feynman1}
\sum_{m} |\Psi_{m}(m,y_{1})|^{2} \kappa^{2} T_{1}^{\mu\nu} \frac{P_{\mu\nu\alpha\beta}^{(m)}}{k^{2}-m^{2}} T_{2}^{\alpha\beta} \Big|_{k^{0} \rightarrow 0}
\end{equation}

The stress energy tensor of the particles that exchange the KK gravitons is $T_{1,2}^{\mu\nu}(\vec{x}) = M_{1,2} \delta(\vec{x}) u^{\mu}u^{\nu}$, where $u^{\mu}$ are the particles' 4-velocities. Taking the particles to be at rest, this gives in momentum space:
\begin{equation}\label{eq:EnergyTensor}
T_{1,2}^{\mu\nu}(\vec{k}) = M_{1,2}  u^{\mu}u^{\nu} = M_{1,2} \delta^{\mu}_{0} \delta^{\nu}_{0}
\end{equation}

This implies that only $P^{(m)}_{0000}$ would contribute. In the limit $k_{0} \rightarrow0$, $P^{(m)}_{0000} = \frac{1}{2}$ for the massless graviton ($m=0$) and $\frac{2}{3}$ for the massive ones ($m>0$). Putting all the pieces together in (\ref{eq:Feynman1}) one obtains:
\begin{equation}\label{eq:GravPotentialK}
U(\vec{k}) = M_{1}M_{2} \kappa^{2} \Bigg[ \frac{|\Psi_{0}(0,y_{1})|^{2}}{2|\vec{k}|^{2}} + \frac{2}{3} \sum_{m} \frac{|\Psi_{m}(m,y_{1})|^{2}}{|\vec{k}|^{2} +m^{2}}   \Bigg]
\end{equation}
where $\Psi_{0}(0,y_{1})$ is normalized to 1. Transforming back to position space and setting $\frac{\kappa^{2}}{8\pi} \equiv G_{N}$, where $G_{N}$ is Newton's constant, we arrive at the following result:
\begin{equation}\label{eq:KKpotential1}
U(r) = \frac{G_{N}M_{1}M_{2}}{r} \Bigg[ 1+\frac{4}{3} \sum_{m} |\Psi_{m}(m,y_{1})|^{2} e^{-m r}  \Bigg]
\end{equation}

In order to evaluate the second term, an analytic expression of $\Psi_{m}(m,y)$ is needed, however, for the warp factor in (\ref{eq:Solutions1}), an analytical solution is not possible. Therefore, I will make some simplifying approximations to evaluate the term. The KK masses are well-approximated by eq. (\ref{eq:KKGravitons}). In addition, $|\Psi_{m}(m,y)|$ does not change appreciably at the IR brane for different $m$, so we can approximate $|\Psi_{m}(m,y_{1})| \hspace{1 mm} \approx |\Psi_{1}(m_{1},y_{1})|$. Although this assumption is somewhat crude, it will not change the results. Performing the sum using (\ref{eq:KKGravitons}), (\ref{eq:KKpotential1}) becomes
\begin{equation}\label{eq:KKPotnetialSum}
U(r) = \frac{G_{N}M_{1}M_{2}}{r} \Bigg[ 1+\frac{4}{3}\frac{|\Psi_{1}(m_{1},y_{1})|^{2}}{e^{\zeta m_{0}r}-1}  \Bigg]
\end{equation}
with $m_{0} \approx 3244$ GeV. One finds that due to the heavy masses of the gravitons, the contribution of the KK modes is highly suppressed to have any relevance to experimental bounds, therefore it can be safely neglected.

The radion's contribution can be calculated in a similar way. Here, the 5D propagator is given by:
\begin{equation}\label{eq:5DRadionPropagator1}
D^{(5)}(x,y;x'y') \equiv \Braket{0|T F(x,y) F^{*}(x',y')|0}
\end{equation}

Decomposing the radion wavefucntion as:
\begin{equation}\label{eq:RadionDecomposition}
F(x,y) = f(y)\sigma(x)
\end{equation}
the 5D propagator for particles localized on the IR brane becomes:
\begin{equation}\label{eq:5DRadionPropagator2}
D^{(5)}(x,y;x'y') = |f(y_{1})|^{2} \Braket{0|T \sigma(x) \sigma(x')|0} =  |f(y_{1})|^{2} D^{(4)}(x;x')
\end{equation}
where $D^{(4)}(x;x')$ is the radion's 4D propagator given by
\begin{equation}\label{eq:radion4DPropagator}
D^{(4)}(x,x') =\int \frac{d^{4}k}{(2\pi)^{4}} \frac{1}{k^{2}-m^{2} +i \epsilon} e^{-i k(x-x')}
\end{equation}

Given that the radion couples to IR matter through the trace of the stress-energy tensor (c.f. \ref{eq:RadionBraneCoupling}):
\begin{equation}\label{eq:RadionIRCoupling}
\frac{1}{\Lambda_{IR}} Tr T_{\mu\nu}
\end{equation}
and that $T_{1,2 \mu}^{\mu} = M_{1,2} \delta_{0}^{\mu} \delta_{0\mu} = 4M_{1,2}$, in the limit $k_{0} \rightarrow 0$ the radion's contribution to gravity gives:
\begin{equation}\label{eq:RadionGravContribution}
U(\vec{k}) =  \frac{16 M_{1} M_{2}}{\Lambda_{IR}^{2}} \frac{1}{|\vec{k}|^{2}+m^{2}} 
\end{equation}
where we have absorbed $|f(y_{1})|$ in the definition of $\Lambda_{IR}$. Transforming (\ref{eq:RadionGravContribution}) to position space and adding all the pieces together, the full modified Newtonian potential is:
\begin{equation}\label{eq:FullNewtonPotential}
U(r) = \frac{G_{N}M_{1}M_{2}}{r} \Bigg[1+\frac{4}{3}\frac{|\Psi_{1}(m_{1},y_{1})|^{2}}{e^{\zeta m_{0}r}-1} +(\frac{4\sqrt{2}}{\kappa \Lambda_{IR}})^{2}  e^{-m_{\sigma} r} \Bigg]
\end{equation}

\section{The $\epsilon$ Dependence of the Radion Coupling}\label{App:Fdep}

Here I investigate the dependence of the radion's wavefunction and coupling on $\epsilon$. Notice that from (\ref{eq:DecayConstant}) we can write the radion's coupling to the IR brane in terms of the \textit{normalized} radion wavefunction $\tilde{F}(y)$ as:

\begin{equation}\label{eq:DecayConstant2}
\Lambda_{IR} = \frac{\sqrt{6}}{\kappa}\frac{1}{\tilde{F}(y_{1})}
\end{equation}

\begin{figure}[!ht]
 \centering
  \includegraphics[height=.3\textheight]{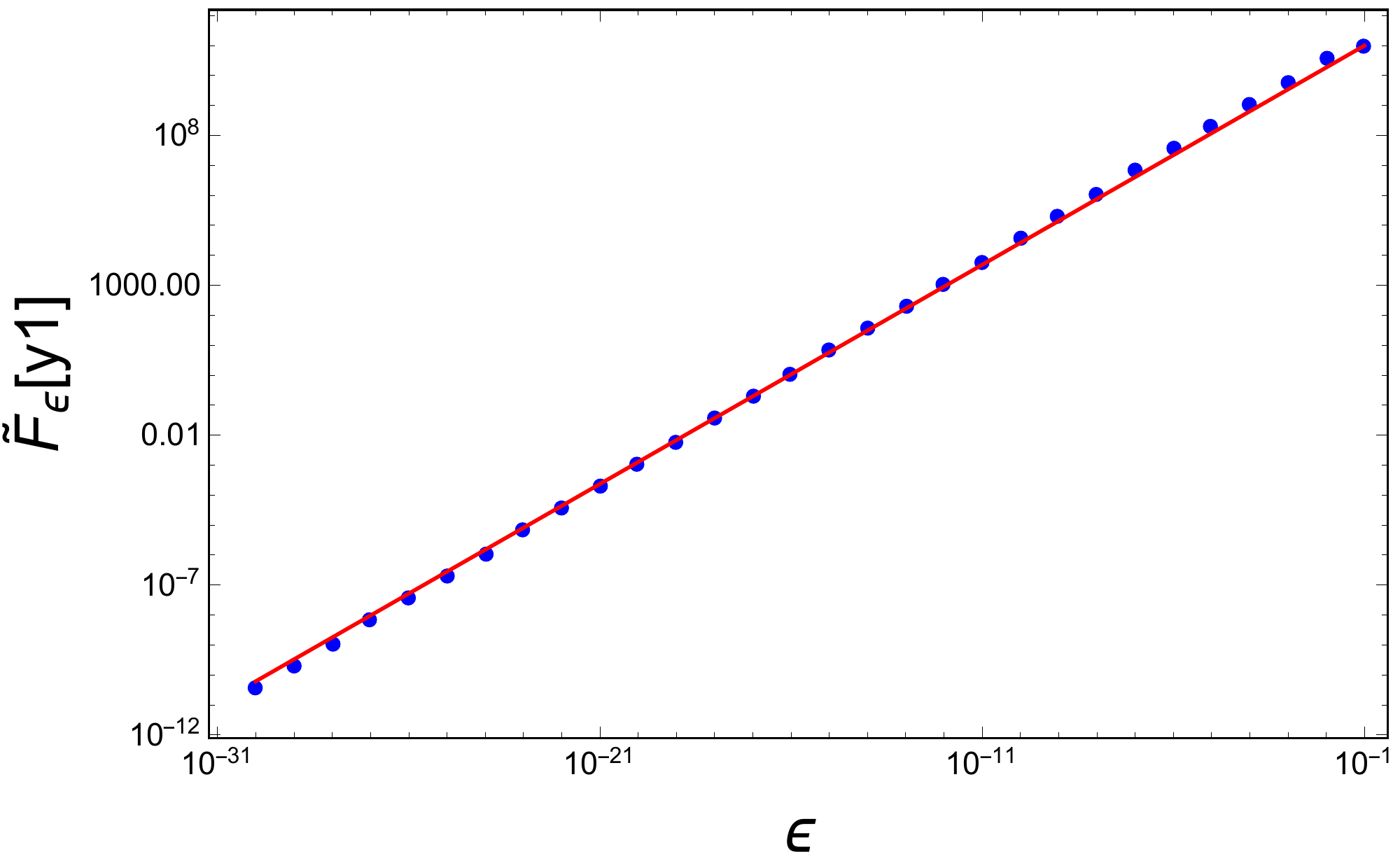}
  \caption{(Blue dots) The dependence of $\tilde{F}(y_{1})$ on $\epsilon$ on a log-log scale. The relationship is fairly linear. (Red line): The fit of this dependence using $\tilde{F}(y_{1}) = a \hspace{1mm} \epsilon^{n}$ for $n \simeq 0.73$.}
  \label{fig:Fdep}
\end{figure}

Since an analytical expression of $\tilde{F}(y)$ is not achievable from (\ref{eq:RadionEOM}), we can investigate its dependence on $\epsilon$ numerically by finding $\tilde{F}_{\epsilon}(y_{1})$ (i.e. on the IR brane) for different values of $\epsilon$.\footnote{Note that in order to do so, one must use the appropriate value of $v_{0}$ that keeps the potential minimum (and hierarchy) fixed, In addition, one must also use the radion mass in (\ref{eq:RadionEOM}) that corresponds to each different value of $\epsilon$, as indicated by Fig. (\ref{fig:alphaDened}).} Fig. (\ref{fig:Fdep}) shows the dependence of $\tilde{F}(y_{1})$ on $\epsilon$ over the range $\epsilon \in [10^{-30},10^{-1}]$ (the blue dots) on a log-log scale. The plot shows a linear relationship between $\log{\tilde{F}(y_{1})}$ and $\log{\epsilon}$, which implied that:

\begin{equation}\label{eq:FVepsilon}
\log{\tilde{F}_{\epsilon}(y_{1})} \propto \log{\epsilon} \hspace{2 mm} \Rightarrow \hspace{2 mm}  \tilde{F}_{\epsilon}(y_{1}) \propto \epsilon^{n}
\end{equation}
for some power $n$. This, together with (\ref{eq:DecayConstant2}), implies that $\Lambda_{IR} \propto \frac{1}{\epsilon^{n}}$. Since the radion's coupling to IR and bulk matter $\sim \frac{1}{\Lambda_{IR}}$ (see Sec. (\ref{Sec:Couplings})), this implies that $g_{\sigma ii} \propto \epsilon^{n}$. This confirms the proportionality found in Fig. (\ref{fig:gVsEpsilon}).
In order to find the exponent $n$, we can find the fit of Fig. (\ref{fig:Fdep}). We find that:

\begin{equation}\label{eq:exponent}
\tilde{F}(y_{1}) \simeq a \hspace{1mm} \epsilon^{0.73}
\end{equation} 
This explains the linear relationship between $\log{g_{\sigma ii}}$ and $\log{\epsilon}$, as is the case in Fig. (\ref{fig:gVsEpsilon}). It also confirms the proportionality of the radion's coupling to $\epsilon$, which means that the coupling becomes smaller for smaller $\epsilon$ (and smaller radion mass), as we saw in our analysis. We should note however, that the value of the exponent remains somewhat unclear and could depend on the choice of other parameters.	Nevertheless, the main point here remains true. This way we can reduce both the mass of the radion and it's coupling, which makes it a viable DM candidate.

\section{Calculation of the Relic Abundance via Freeze-in}\label{App:RelicAbundance}
Here I calculate the radion abundance following the treatment described in \cite{Hall:2009bx}. For $\gamma p \rightarrow \sigma p$, the Boltzmann equation reads:

\begin{dmath}\label{eq:Boltzmann1}
\dot{n}_{\sigma} + 3H n_{\sigma} = \int d\Pi_{\sigma}d\Pi_{\gamma}d\Pi_{p_{i}}d\Pi_{p_{f}} (2\pi)^{4} \delta^{4}(p_{1}+p_{2}-p_{3}-p_{4}) \times \Big[ |M_{p \gamma \rightarrow \sigma p}|^{2} f_{p_{i}} f_{\gamma} (1-f_{p_{f}})(1+f_{\sigma}) -  |M_{\sigma p \rightarrow p\gamma}|^{2}f_{\sigma}f_{p_{i}}(1+f_{\gamma})(1-f_{p_{f}}) \Big]
\end{dmath}
where $d\Pi_{i}$ is the Lorentz-invariant phase space and $p_{i,f}$ refer to the initial and final protons respectively. Neglecting the blocking/stimulation factors, and assuming CPT invariance and that the radion's initial abundance is negligible, the equation simplifies to:

\begin{dmath}\label{eq:Boltzmann1}
\dot{n}_{\sigma} + 3H n_{\sigma} = \int d\Pi_{\sigma}d\Pi_{\gamma}d\Pi_{p_{i}}d\Pi_{p_{f}} (2\pi)^{4} \delta^{4}(p_{1}+p_{2}-p_{3}-p_{4}) \times |M_{p\gamma \rightarrow \sigma p}|^{2} f_{p_{i}} f_{\gamma}\\
 \equiv (n_{\sigma}^{EQ})^{2} \braket{\sigma|\vec{v}|}
\end{dmath}
where $(n_{\sigma}^{EQ})$ is the equilibrium radion number density $= \zeta(3) T^{3}/\pi^{2}$ and $\braket{\sigma|\vec{v}|}$ is the thermally-averaged annihilation cross section, which can be written in the Lorentz-invariant form \cite{Gondolo:1990dk}:

\begin{equation}\label{eq:ThermallyAveragedXSection}
\braket{\sigma|\vec{v}|} = \frac{1}{16m^{2} T^{3} K_{2} (\frac{m}{T})} \int_{m^{2}}^{\infty} ds (s-m^{2})^{3/2}  \sigma(s) K_{1} \Big(\frac{\sqrt{s}}{T} \Big)
\end{equation}
where $K_{1,2}(x)$ are the Modified Bessel functions of the first and second kind, and $m$ is the mass of the proton. Using $Y= n/S$, the L.H.S. of the Boltzmann equation becomes:

\begin{equation}\label{eq:LHS1}
\dot{Y}S + 3H S Y
\end{equation} 
Since we are assuming negligible initial abundance, $Y \approx 0$ and using $\dot{T} \approx -H T$, the L.H.S. simplifies to:

\begin{equation}\label{eq:LHS2}
H T S \frac{dY}{dT}
\end{equation} 
Substituting (\ref{eq:XSection}) in (\ref{eq:ThermallyAveragedXSection}) and using the explicit expressions for $H$ and $S$ given by:

\begin{equation}\label{eq:S}
S = \frac{2 \pi^{2} g_{*s}T^{3}}{45}
\end{equation}

\begin{equation}\label{eq:H}
H = \sqrt{\frac{g_{*}}{90}}\pi \frac{T^{2}}{M_{Pl}}
\end{equation}
the Boltzmann equation becomes:
\begin{equation}\label{eq:Boltzmann2}
\frac{dY}{dT} = \sqrt{\frac{90}{g_{*}}} \frac{45 \zeta^{2}(3) a e^{2} g_{\sigma NN}^{2} M_{Pl}}{128 \pi^{8} g_{*s}m^{2} T^{3} K_{2}(\frac{m}{T})} \int_{m^{2}}^{\infty}ds \frac{(s-m^{2})^{3/2}}{s} K_{1} \Big(\frac{\sqrt{s}}{T} \Big)
\end{equation}
where $a=2s+1$ is the proton degeneracy. The integral gives:

\begin{equation}\label{eq:IntegralResult}
\frac{3}{8}\sqrt{\pi} m^{3} \MeijerG{3,}{0}{1,}{3}{1}{-\frac{3}{2},-\frac{1}{2},\frac{1}{2}}{\frac{m^{2}}{4T^{2}}}
\end{equation}
 
Defining $x \equiv \frac{m}{T}$, (\ref{eq:Boltzmann2}) becomes:

\begin{dmath}\label{eq:Boltzmann3}
\frac{dY}{dx} = \sqrt{\frac{90 \pi}{g_{*}}} \frac{135 \zeta^{2}(3) a e^{2} g_{\sigma NN}^{2} M_{Pl}}{1024 \pi^{8} g_{*s}m^{}} \frac{x}{K_{2}(x)}  \MeijerG{3,}{0}{1,}{3}{1}{-\frac{3}{2},-\frac{1}{2},\frac{1}{2}}{\frac{x^{2}}{4}}\\
\approx 1.7171 \times 10^{14}\frac{g_{\sigma NN}^{2}}{g_{*s}\sqrt{g_{*}}} \frac{x}{K_{2}(x)}  \MeijerG{3,}{0}{1,}{3}{1}{-\frac{3}{2},-\frac{1}{2},\frac{1}{2}}{\frac{x^{2}}{4}}
\end{dmath}

This equation can be integrated numerically to give the final yield:

\begin{equation}\label{eq:FinalYield}
Y_{\infty} \approx 1.7171 \times 10^{14}\frac{g_{\sigma NN}^{2}}{g_{*s}\sqrt{g_{*}}}   \int_{0}^{\infty} dx \frac{x}{K_{2}(x)}  \MeijerG{3,}{0}{1,}{3}{1}{-\frac{3}{2},-\frac{1}{2},\frac{1}{2}}{\frac{x^{2}}{4}}
\end{equation}

from which the radion relic abundance can be found as:

\begin{equation}\label{eq:FinalRelic}
\Omega h^{2} \approx 3.325 \times 10^{27} \frac{m_{\sigma}}{M_{Pl}} Y_{\infty}
\end{equation}

\end{document}